\documentclass[iop,apjl, twocolappendix]{emulateapj}   

\pdfoutput=1

\usepackage{graphicx}
\usepackage{epsfig}
\usepackage{amsmath}
\usepackage{subfigure} 
\usepackage{hyperref}
\usepackage{amssymb}
\usepackage{apjfonts}
\usepackage{amsmath}
\usepackage{latexsym}
\usepackage{color}
\usepackage[usenames,dvipsnames,svgnames,table]{xcolor}
\usepackage{hyperref}
\usepackage[all]{hypcap}    
\hypersetup{
    colorlinks,
    citecolor=blue,
    filecolor=blue,
    linkcolor=blue,
    urlcolor=blue
}

\def\kms{\ifmmode{~{\rm km~s^{-1}}}\else{~km s$^{-1}$}\fi}
\def\cc{\ifmmode{~{\rm cm^{-3}}}\else{~cm$^{-3}$}\fi}
\def\fesc{\ifmmode{{f_{esc}}}\else{$f_{\rm esc}$}\fi}
\def\fstar{\ifmmode{{f_\star}}\else{$f_\star$}\fi}
\def\lsim{\lower0.3em\hbox{$\,\buildrel <\over\sim\,$}}
\def\gsim{\lower0.3em\hbox{$\,\buildrel >\over\sim\,$}}

\def\enzo{{\sc Enzo}}

\def\Ms{\ifmmode{~{\rm M_\odot}}\else{M$_\odot$}\fi}
\def\Zs{\ifmmode{~{\rm Z_\odot}}\else{Z$_\odot$}\fi}
\def\h2{H$_2$}

\newcommand{\vvec}[1]{\mathbf{#1}}
\newcommand{\hhat}[1]{\widehat{#1}}
\newcommand{\astc}{^\ast}

\shorttitle{Magnetic length scales in galaxy clusters} 
\shortauthors{H. Egan et al.}

\begin{document}\title{Length Scales and Turbulent Properties of Magnetic Fields
  in Simulated Galaxy Clusters}

\author{
  Hilary Egan\altaffilmark{1},
  Brian W. O'Shea\altaffilmark{2},
  Eric Hallman\altaffilmark{3},
  Jack Burns\altaffilmark{1},
  Hao Xu\altaffilmark{4},
  David Collins\altaffilmark{5},
  Hui Li\altaffilmark{6}, and
  Michael L. Norman\altaffilmark{4}
}

 \affil{$^{1}${Center for Astrophysics and Space Astronomy,
   Department of Astrophysical and Planetary Sciences, University of
   Colorado, Boulder, CO 80309, USA 
 \href{mailto:Hilary.Egan@colorado.edu}{Hilary.Egan@colorado.edu}}}


\affil{$^{2}${Department of Computational Mathematics, Science and
    Engineering, Department of Physics and Astronomy, and National
    Superconducting Cyclotron Laboratory, Michigan State
    University, East Lansing, MI 48824, USA
    \href{mailto:oshea@msu.edu}{oshea@msu.edu}}}

\affil{$^{3}${Tech-X Corporation, Boulder, CO 80303, USA}}

\affil{$^{4}${University of California at San Diego, La Jolla, CA
    92093, USA}}

\affil{$^{5}${Florida State University,  Tallahassee, FL 32306, USA} }

\affil{$^{6}${Los Alamos National Laboratory, Los Alamos, NM 87545, USA}}

\label{firstpage}

\begin{abstract}
Additional physics beyond standard hydrodynamics is needed to fully model the
intracluster medium (ICM); however, as we move to more sophisticated models, it
is critical to consider the role of magnetic fields and the way the fluid
approximation breaks down. This paper represents a first step towards developing
a self-consistent model of the ICM by characterizing the statistical properties
of magnetic fields in cosmological simulations of galaxy clusters. We find that
plasma conditions are largely homogeneous across a range of cluster masses and
relaxation states. We also find that the magnetic field length scales are 
resolution dependent and not based on any specific physical process. Energy 
transfer mechanisms and scales are also identified, and imply the existence
of small scale dynamo action. The scales of the small scale dynamo are 
resolution-limited and are likely to be driven by numerical resistivity and viscosity. 
\end{abstract}

\keywords{galaxy clusters: plasma -- galaxy clusters: intracluster medium -- magnetohydrodynamics}

\section{Introduction}

Simulations of galaxy clusters have made rapid advances in the past
few years, and match observations of galaxy clusters in a wide variety
of ways -- generally, observationally determined density, temperature,
and entropy profiles of clusters are reproduced in simulations, at
least outside the central regions \citep[see, e.g.,][]{hall06, nagai,
borg_krav}.  This matching indicates that the intracluster
medium (ICM) properties, at least in terms of the integrated observables
in the cluster volume, are largely driven by the gravitational potential
of the dark matter halo in which the ICM plasma resides, and that 
simple hydrodynamics and gravity can account for the general behavior
of the intracluster plasma \citep{2010ApJ...721.1105B}.

There are some observational incongruities regarding the intracluster
medium, however -- in terms of bulk quantities,  simulations do a
poor job of matching the properties of the intracluster medium in
cluster cores, particularly in cool-core clusters
\citep[e.g.,][]{2013ApJ...763...38S}.  Also, there are some
observational features, such as cluster cold fronts \citep[see,
e.g.,][]{2007PhR...443....1M,2002AstL...28..495V}, that are difficult
to explain with our standard models and require more sophisticated
plasma physics to understand
\citep[e.g.,][]{2015ApJ...798...90Z, 2015arXiv150606429W}.

More broadly, it is clear that additional physics is needed to model
the intracluster medium (ICM): synchrotron radiation from radio relics
and radio halos indicate that there is acceleration of charged particles
to relativistic speeds by some combination of the first- and second-order
Fermi processes \citep{clarke_enss,giacintucci,vanweeren, 2013ApJ...765...21S,
2014ApJ...793...80D}.  This also requires the presence of magnetic fields.
Similarly, other observations, such as Faraday rotation measures of background
light passing through the ICM \citep{2010A&A...513A..30B,2004A&A...424..429M},
show that magnetic fields are ubiquitous in clusters, both in the cluster core
and in the outskirts of clusters \citep{2002ARA&A..40..319C,clarke_enss,
1992ApJ...388L..49B}. While magnetic fields are typically dynamically
unimportant (plasma $\beta$ values estimated from observation are $\gg 1$
in essentially all of the cluster outside of AGN jets and, perhaps, jet-driven bubbles), they
are essential to reproducing these types of observations.


Estimates of the Reynolds number ($Re$), magnetic Reynolds number ($R_M$),
and Prandtl number ($Pr$) suggest that a small scale dynamo process will
work to amplify a small seed magnetic field, possibly to levels observed in
the ICM \citep{2006MNRAS.366.1437S}.  As has been argued in the recent work by
\citet{2014ApJ...782...21M, 2015ApJ...800...60M}, the plasma in galaxy clusters
has an extremely high Reynolds number ($Re\equiv UL/\nu$, having effectively
zero viscosity), and thus must develop Kolmogorov-like turbulence very rapidly.
The ICM also satisfies $\eta \ll \nu$ \citep{2005PhR...417....1B, 
2006MNRAS.366.1437S}, leading to large Prandtl numbers ($Pr\equiv \nu/\eta \gg 1$)
and magnetic Reynolds numbers ($R_M \equiv UL/\eta \gg Re$). 

Limitations in current simulation capabilities mean that it is infeasible
to simulate fluids with such large Reynolds and Prandtl numbers, as both
ideal viscosity and resisitivity are much smaller than numerical viscosity
and resisitivity. Furthermore, the highest Reynolds numbers are only 
achieved in turbulent box simulations; cosmological scale simulations are 
further limited ($Re\sim500-1000$). These conditions drastically change 
the nature of any small scale dynamo, if indeed they are even capable of 
exciting one.


There are various physical processes that have been modeled in
clusters in recent years -- thermal conduction, viscosity, and subgrid
turbulence -- that are typically treated as independent \citep[e.g.,
][]{2013ApJ...778..152S, 2009MNRAS.395.2210B}, but are all critically
dependent on the behavior of the magnetic field and its interaction
with the plasma at both observable scales and below.  As a result, a
truly self-consistent model of the intracluster plasma must treat all
of these phenomena, and others as well, as manifestations of local
plasma properties. The magnetic field evolution will thus depend on the
the effects of the subgrid properties and the subgrid properties must by
influenced by the local magnetic field.  

Making such a model is a significant theoretical challenge, and due to
practical constraints it is beyond the scope of cosmological
simulations of galaxy cluster formation and evolution -- simulations
that resolve all relevant scales in galaxy clusters, from cosmological
scales down to the turbulent dissipation scale, are computationally
unfeasible.  However, an important first step is to characterize the
statistical properties of magnetic fields in cosmological simulations
of galaxy clusters, which will serve to motivate more sophisticated
plasma modeling at a later date. We must examine how and if amplification
is taking place and what signatures it is imparting on the magnetic field,
as well as the numeric and model-dependent resistive features. Although
such processes have been examined in isolation in the form of turbulent
box simulations, it also necessary to compare to full cosmological simulations
with the effects of  merger driven turbulence and strong large scale
density gradients. 

Our goal is to analyze the magnetic field amplification signatures of
a set of 12 high-resolution cosmological magnetohydrodynamical (MHD) 
simulations of galaxy clusters from \citet{2009ApJ...698L..14X,
2010ApJ...725.2152X,2011ApJ...739...77X,2012ApJ...759...40X},
as they relate to the numerical effects arising from resolution limitations.
We generally focus on the statistical properties of the cluster turbulence and
magnetic fields as well as the differences between clusters
that are morphologically relaxed in X-ray observations, and those that
are unrelaxed (i.e., that display evidence of recent mergers).

This paper is organized as follows.  In Section \ref{sec:methods} we
discuss the simulation setup and the general characteristics of our
cluster sample. In Section \ref{sec:results} we present our main
results including a selection of global cluster properties
(Sec. \ref{sec:similarity}), identification of the scale lengths
present in the plasma (\ref{sec:autocor}), and energetic properties
of the plasma (\ref{sec:struc}, \ref{sec:shellshell}). We discuss these
results in the larger context of ICM plasma simulations in Section
\ref{sec:discussion}.  Finally, Section \ref{sec:summary} contains a
summary of our findings.

\section{Methods}
\label{sec:methods}

\subsection{Simulation}

The simulations we analyze were previously studied in
\citet{2012ApJ...759...40X}. They were run using the cosmological code \enzo\
\citep{2014ApJS..211...19B}, with the \enzo+MHD module \citep{collins_2010}. The
cosmological model chosen was a $\Lambda$CDM model with cosmological parameters
$h = 0.73$, $\Omega_m=0.27$, $\Omega_b=0.044$, $\Omega_\Lambda=0.73$, and 
$\sigma_8 = 0.77$.

To simulate the clusters, eleven boxes with different random initial seeds were
run. The simulations were evolved from $z=30$ to $z=0$. An adiabatic equation of
state was used ($\gamma=5/3$), and no additional physics such as radiative
cooling or star formation was used. The magnetic field was injected using
the magnetic tower model \citep{2008ApJ...681L..61X,2006ApJ...643...92L}
at $z=3$. The tower model is designed to represent the large scale effects
of a magnetic energy dominated AGN jet, and the total energy injected 
($\sim6\times10^{59}~$erg over $\sim 30~$Myr) is similar to the energy 
input from a moderately powerful AGN. One cluster was run twice with
magnetic field being injected in a different protocluster during each run.  

Each cluster was simulated in a $(256~h^{-1} $Mpc$)^3$ box with a
$128^3$ root grid.  Two levels
of static nested grids were used in the Lagrangian region where the
cluster is formed such that the dark matter particle resolution is
$1.07\times10^7 M_\odot$. A maximum of eight levels of refinement were also
applied in a ($\sim 50$ Mpc)$^3$ box centered on the location the galaxy
cluster forms which combine to give a maximum resolution of $7.81~kpc~h^{-1}$. 
One cluster (marked in Table~\ref{tab:clusters}) was simulated again with an
additional level of refinement for a maximum resolution of $3.91~$kpc$~h^{-1}$
The refinement is based
on the baryon and dark matter overdensity during the course of cluster
formation, and after the magnetic field is injected, all regions where
the magnetic field strength is greater than $5\times10^{-8}$~G are refined
to the highest level. On average, $90-95\%$ of the cells within the virial
radius of the cluster are refined at the highest level. 
\subsection{Cluster Sample}

From our simulations we gather a sample of 12 total clusters, with 2 clusters
having the same initial conditions but different magnetic field injection 
sites. The clusters cover a range of masses from
($1.44\times10^{14}-1.8\times10^{15}~M_\odot$). We separate
the sample into two groups (relaxed and unrelaxed) based on whether they have
accumulated more than half of their final mass by a redshift of $z=0.5$ as
described by \citet{2011ApJ...739...77X}.
A full summary of the cluster properties is given in Table~\ref{tab:clusters}. 
A star indicates that the clusters are the same final cluster, but the
magnetic fields were injected into separate proto-clusters.

\begin{table} 
  \caption{Cluster Sample}
  \centering
  \begin{tabular}{llll}
    \hline               
    $R_{200}$ (Mpc) & $M_{200}$ ($M_\odot\times 10^{14}$) & Temperature (KeV) & Relaxation\\ \hline     
    4.08 & 18.01 & 9.04 & R $^\dagger$\\
    3.45 & 11.95 & 6.80 & R \\
    3.01 & 7.53 & 5.38 & R \\
    3.00 & 5.94 & 3.80 & R \\
    2.34 & 2.97 & 2.74 & R \\
    1.81 & 1.44 & 1.65 & R \\
    2.74 & 7.52 & 5.55 & U* \\
    2.34 & 7.18 & 5.51 & U* \\
    3.30 & 8.75 & 5.88 & U \\
    2.68 & 6.99 & 5.43 & U \\
    3.16 & 8.44 & 5.75 & U \\
    2.99 & 5.93 & 4.39 & U \\\hline
  \end{tabular}
  \label{tab:clusters}
  \tablecomments{Cluster properties for the full sample of
    clusters. $R_{200}$ and
  $M_{200}$ are calculated with respect to the critical density using
  spheres centered via iterative center of mass calculations. The temperature
  is calculated using the virial mass-temperature relationship ($k_BT=(8.2 keV)(M_{200}/10^{15}M_\odot)^{2/3}$, at $z=0$ \citep{2005RvMP...77..207V}), and relaxation
  is defined by having more or less than half the final mass of the cluster
  accumulated by $z=0.5$. Asterisks indicate that the two clusters
  are the same final cluster, but the magnetic field was injected into separate
  proto-clusters. $\dagger$ indicates that this cluster was resimulated at a
  higher resolution.}  
\end{table}

\begin{figure*}
  \centering
          \includegraphics[width=0.45\textwidth]{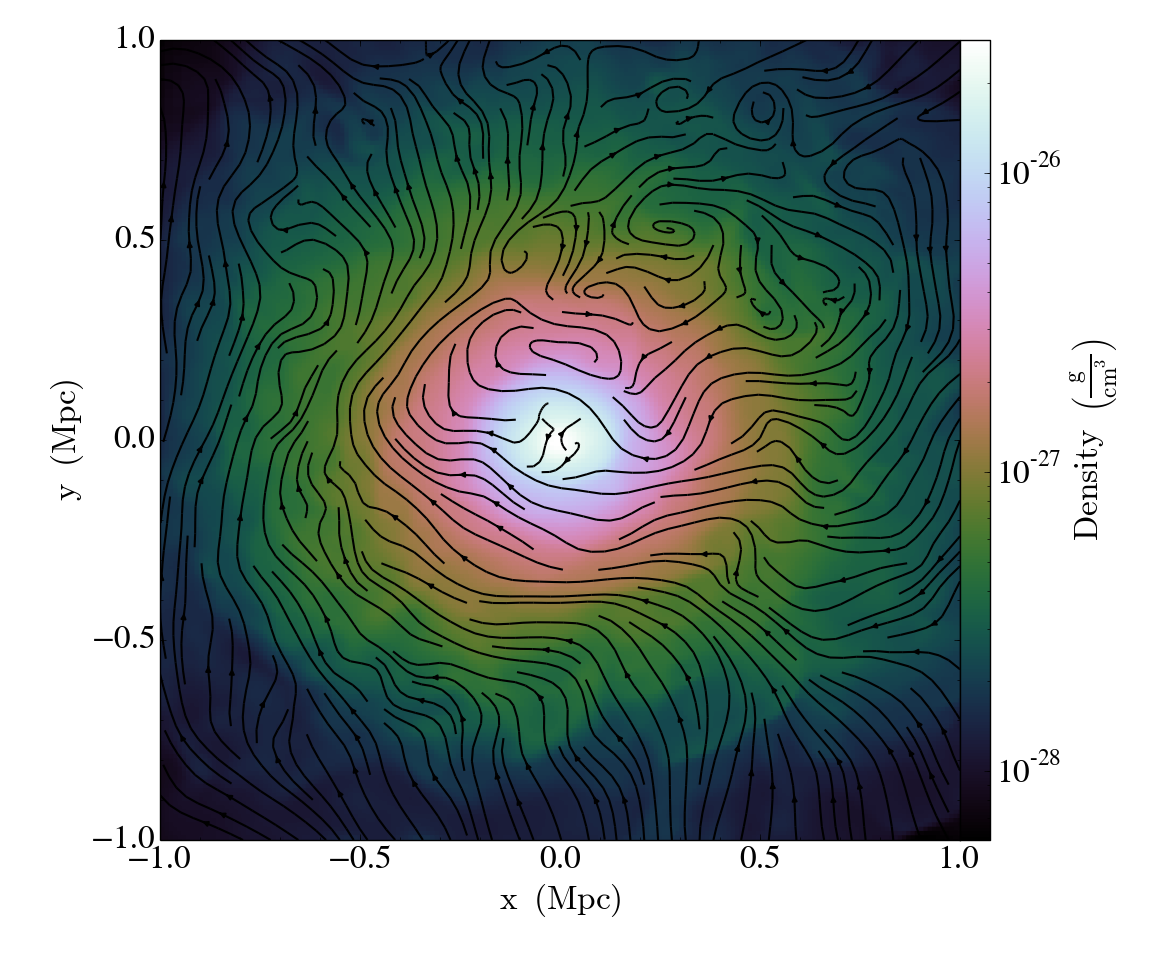}
          \includegraphics[width=0.45\textwidth]{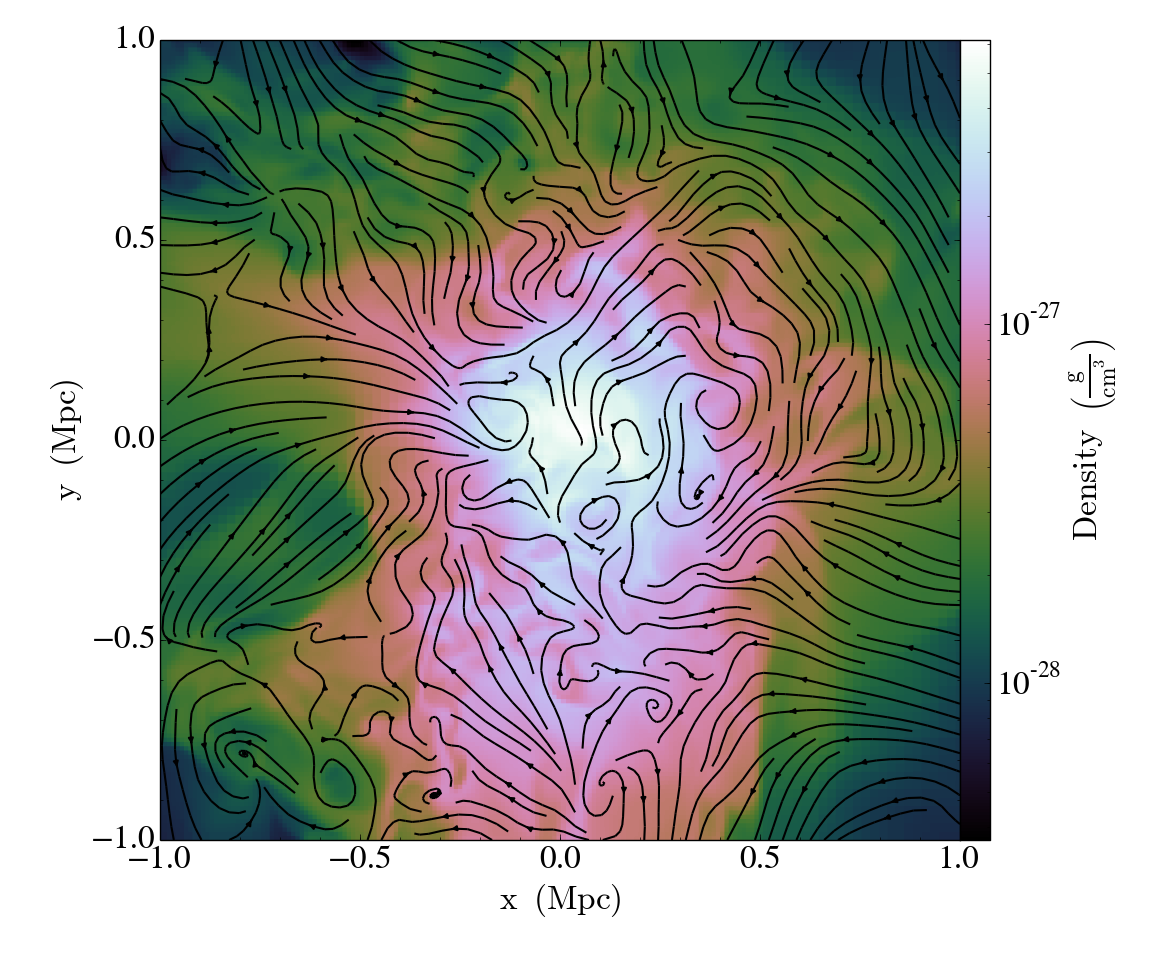}
          \includegraphics[width=0.45\textwidth]{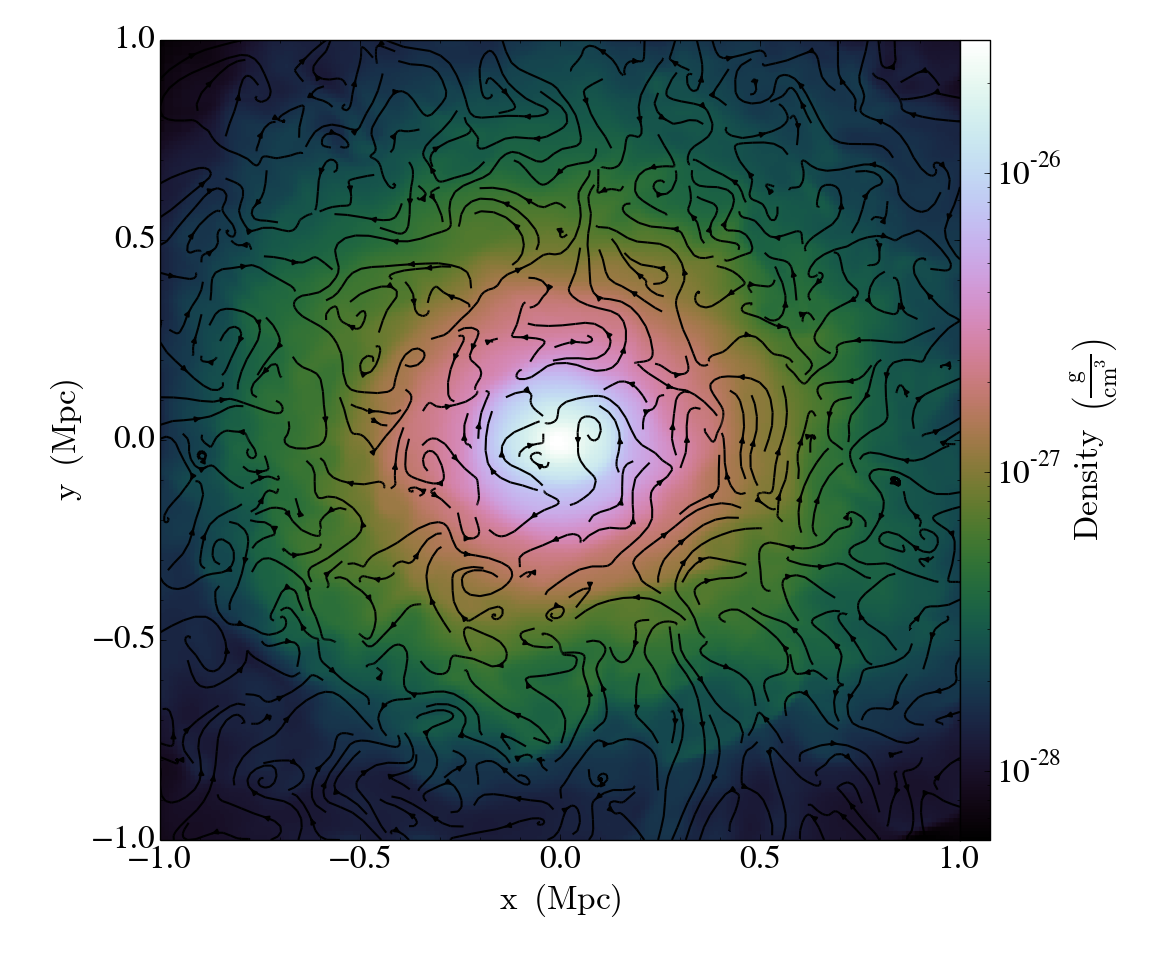}
          \includegraphics[width=0.45\textwidth]{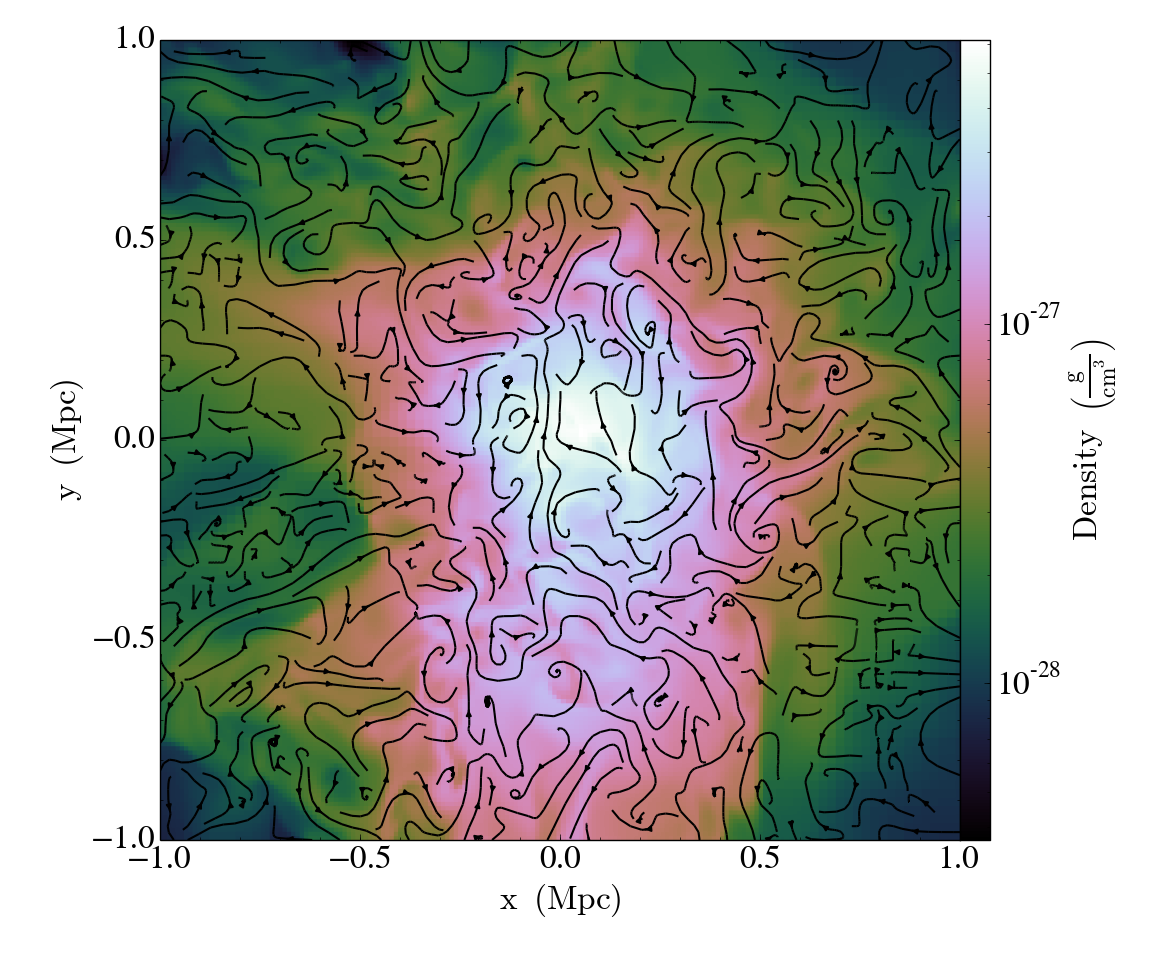}\\
  \caption{Streamlines for velocity (\emph{top}) and magnetic field (\emph{bottom}) 
           plotted over density for a relaxed (\emph{left}) and an unrelaxed
           (\emph{right}) cluster. Note that while the velocity substructure
           shows a clear difference between relaxation states, the magnetic
           field structure appears comparably tangled in both cases.}
  \label{fig:stream}
  \vspace{8mm}
This figure, and some of the following analysis was completed with the aid
of open source, volumetric data analysis package, yt \citep{2011ApJS..192....9T}.
\end{figure*}

Figure \ref{fig:stream} shows a sample of two clusters, a relaxed and an
unrelaxed cluster, with velocity and magnetic field streamlines for a
thin slice overplotted. 

\section{Results}
  \label{sec:results}

  \subsection{Global Magntic Field Properties}
\label{sec:similarity}
  
\subsubsection{Profiles}
\begin{figure}
  \centering
  \includegraphics[width=0.45\textwidth]{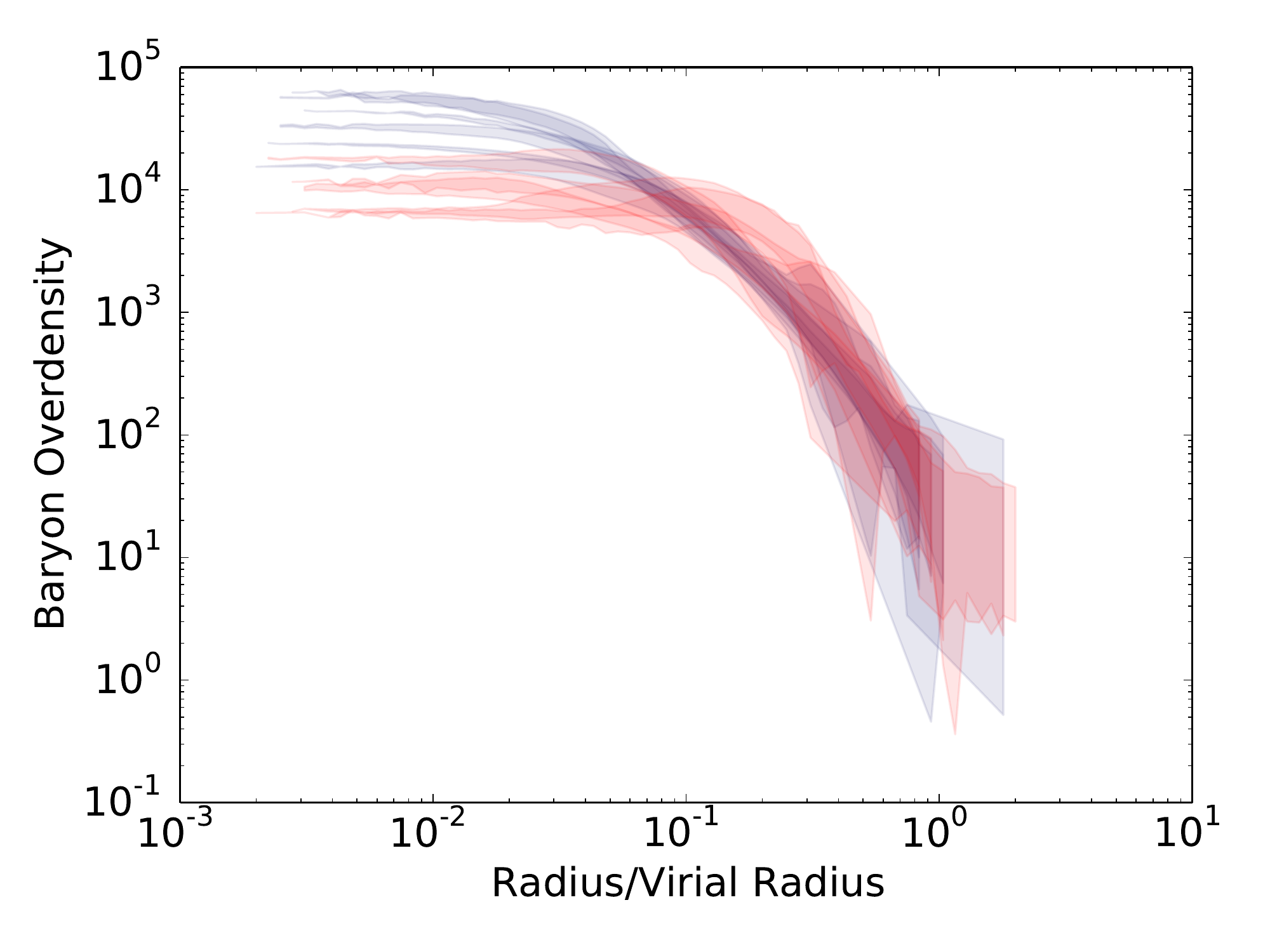}
  \caption{Volume-averaged baryon overdensity measured in spherical shells as a
           function of radius for all clusters. Color indicates relaxation
           state, where \emph{red} is unrelaxed and \emph{blue} is relaxed.
           Shaded regions show one standard deviation above and below the mean.
           This figure may be used as a point of reference for where in the
           cluster typical baryon overdensities can be found, as used in 
           Figures~\ref{fig:prof_mag_std} and
           ~\ref{fig:phase_plasma_beta}.}
\label{fig:prof_dens_std}
\end{figure}

One inherent difficulty in simulating clusters is the extreme range in
scales from the centers of clusters to their outskirts. Baryon overdensity
typically changes over several orders of magnitude (as seen in Figure~
\ref{fig:prof_dens_std}), which can influence many properties that affect
the turbulent state of the plasma. As such, for much of the following analysis
we break down the quantities by baryon overdensity or radius from the center
of the cluster.

\begin{figure*}
  \centering
  \includegraphics[width=0.45\textwidth]{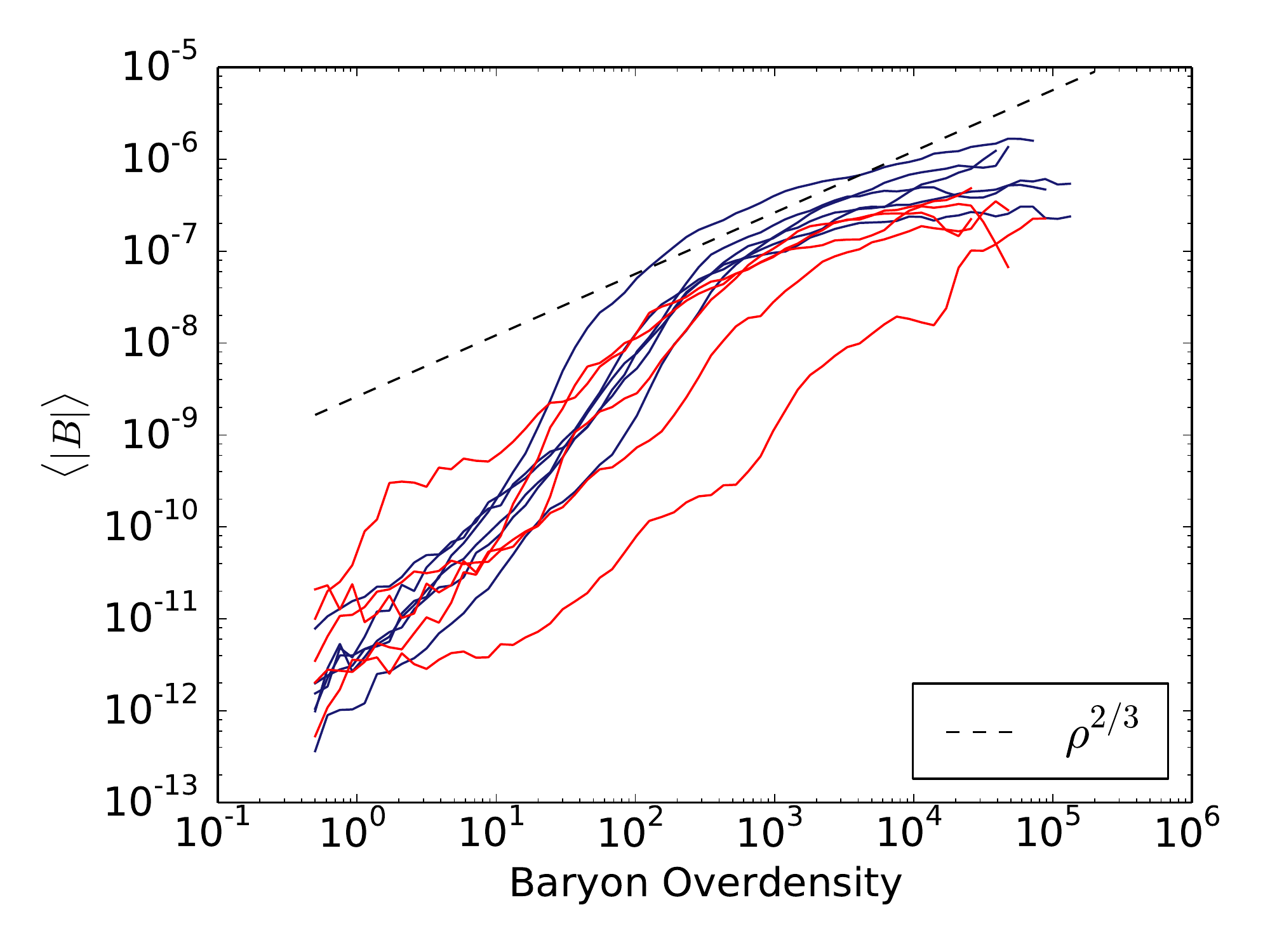}
  \includegraphics[width=0.45\textwidth]{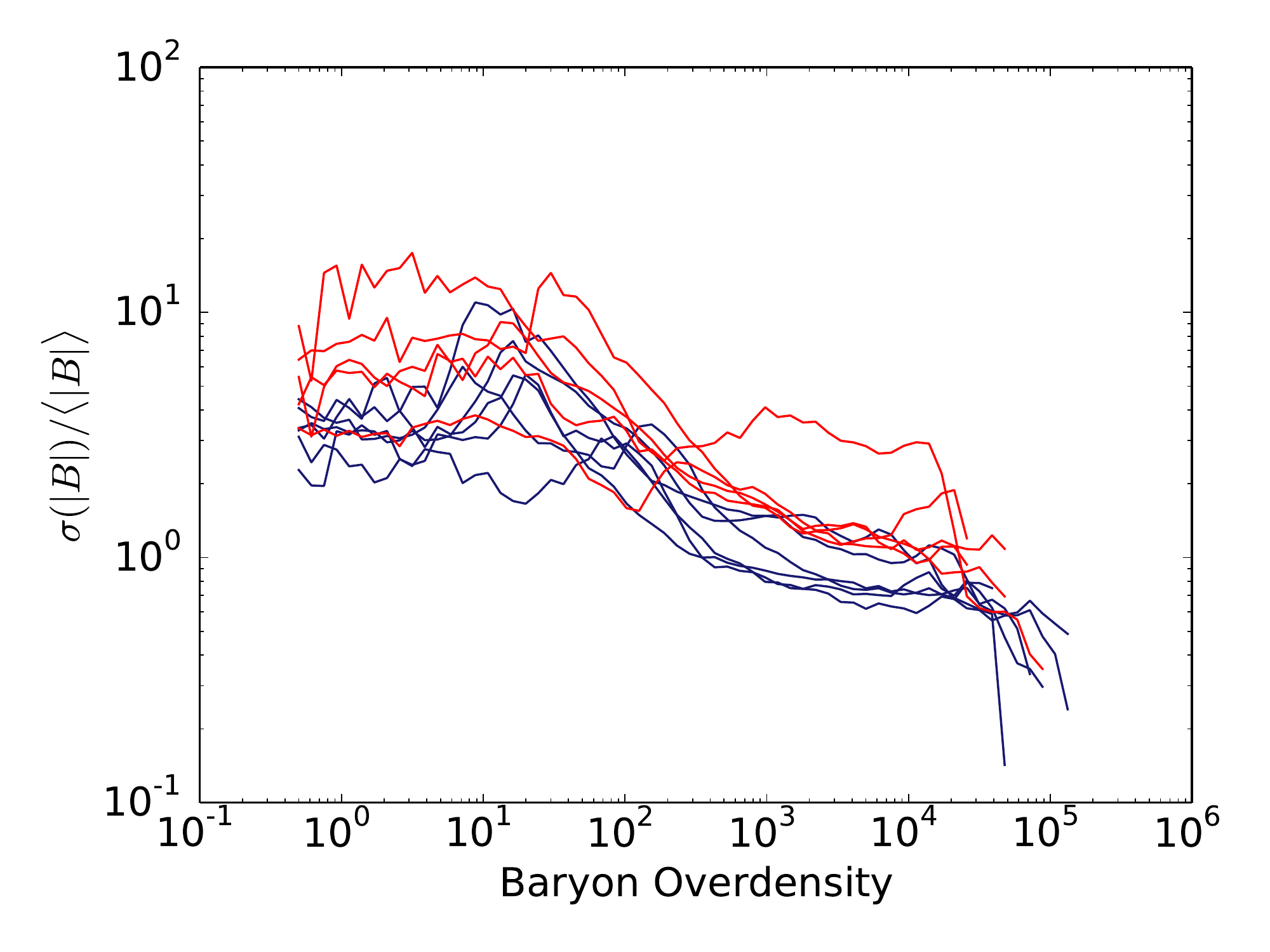}
  \caption{Volume-averaged magnetic field magnitude versus baryon
           overdensity for relaxed and unrelaxed clusters. Red
           signifies unrelaxed clusters while blue indicates relaxed,
           with one line per cluster. \emph{Left:} Mean value of
           magnetic field magnitude, \emph{Right:}  variance
           of the magnetic field magnitude at a given baryon overdensity
           divided by the mean value of magnetic field magnitude.
           The dashed line in the left figure shows $B\propto\rho^{2/3}$.}
\label{fig:prof_mag_std}
\end{figure*}

As shown in Figure~\ref{fig:prof_mag_std}, the magnetic field is not 
amplified uniformly throughout the cluster. If flux-freezing conditions
were completely satisfied and the collapse were completely spherical,
 the magnetic the magnetic field magnitude would be 
proportional to  $\rho^{2/3}$; however, this is only true at baryon
overdensities greater than $10^2-10^3$ or radii less than roughly half the 
virial radius of the cluster. This behavior is consistent across relaxed
and unrelaxed clusters, although there is one unrelaxed cluster that does
not get amplified to the $\rho^{2/3}$ level at all. This is the least 
massive cluster, and it is likely that it was injected with too large
of a magnetic field for the size of the cluster. 

Figure~\ref{fig:prof_mag_std} shows the standard deviation around the
mean of the magnetic field magnitude shown in Figure~\ref{fig:prof_mag_std}.
In general the scatter in quantities rises as the overdensity
falls, indicating that  conditions vary more widely at cluster
outskirts. Additionally, the variance is generally larger for the
unrelaxed clusters.

\begin{figure*}
  \centering
  \includegraphics[width=0.95\textwidth]{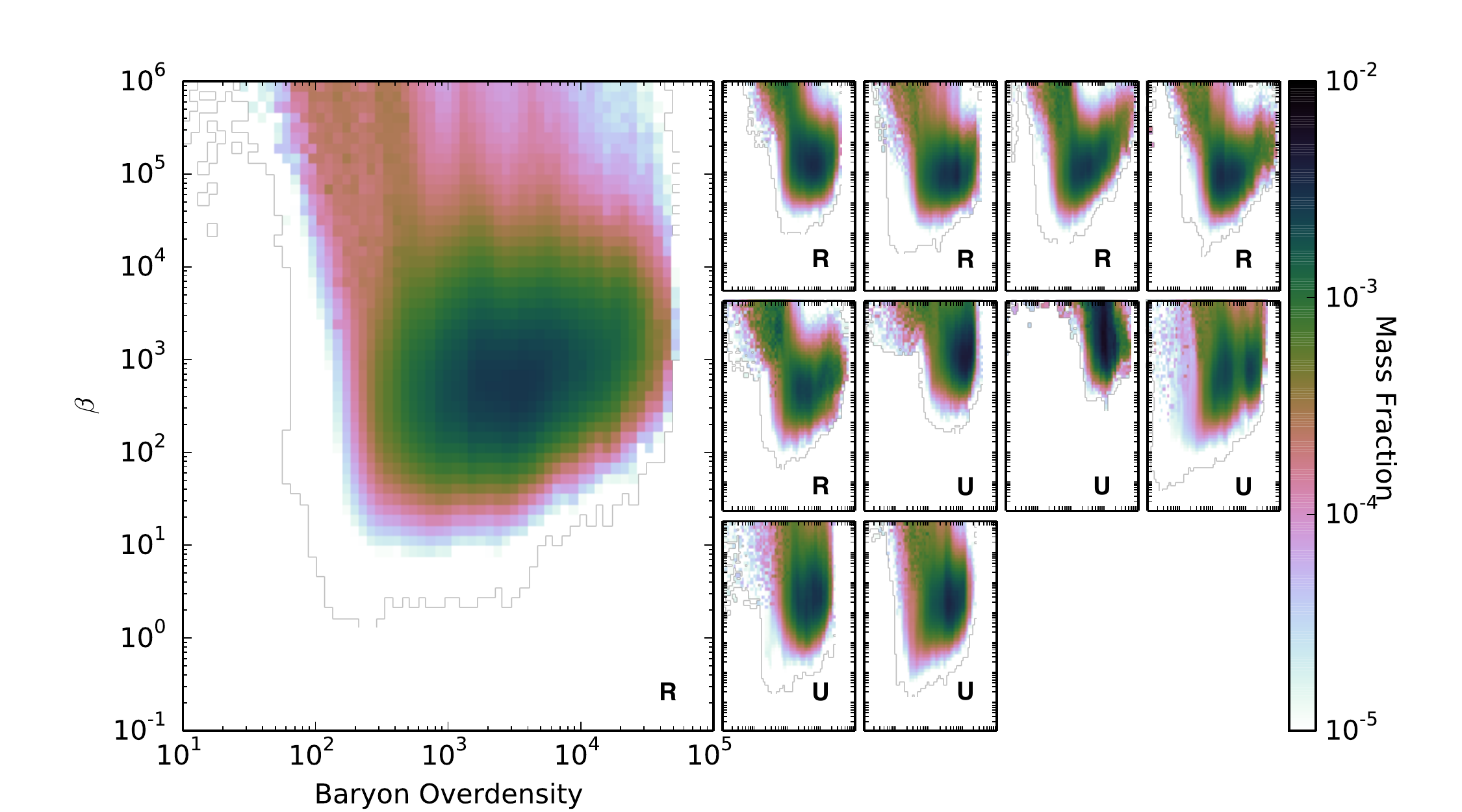}
  \caption{2D gas mass-weighted histogram of plasma beta vs.\ baryon overdensity.
           Each panel is a separate cluster with the letters indicating
           relaxed (\emph{R}) or unrelaxed (\emph{U}). All panels have 
           the same range of baryon overdensity and plasma beta, and
           the total gas mass is normalized to 1 for direct comparison.}
\label{fig:phase_plasma_beta}
\end{figure*}

Plasma $\beta$ (defined as $\beta=\frac{P_{thermal}}{P_{mag}}=
\frac{nk_BT}{B^2/(2\mu_0)}$) measures the dynamical importance of the magnetic
field pressure. Although these values are typically much greater than 1,
the exact distribution varies
substantially from cluster to cluster. Figure~\ref{fig:phase_plasma_beta}
shows a selection of 2D gas mass-weighted histograms of $\beta$ vs
baryon overdensity, with one panel per cluster. In general these
distributions are centered around plasma betas of $\sim10^2-10^3$ and
baryon overdensities of $\sim10^3-10^4$; however, a few clusters have
substantial gas mass in tails extending to much higher plasma betas. These
can likely be attributed to infalling clumps of gas that have not yet
been magnetized.

\subsection{Scale Lengths}

In this section we discuss a variety of scale lengths present in
the plasma (Jeans length, characteristic resistive length, and
autocorrelation length) as compared to the cell size. The comparison
of these scale lengths with the cell size can give insight into the
interaction of small scale turbulence with  numerical effects due
to finite resolution.

\subsubsection{Jeans Length}
\begin{figure*}
  \centering
  \includegraphics[width=0.45\textwidth]{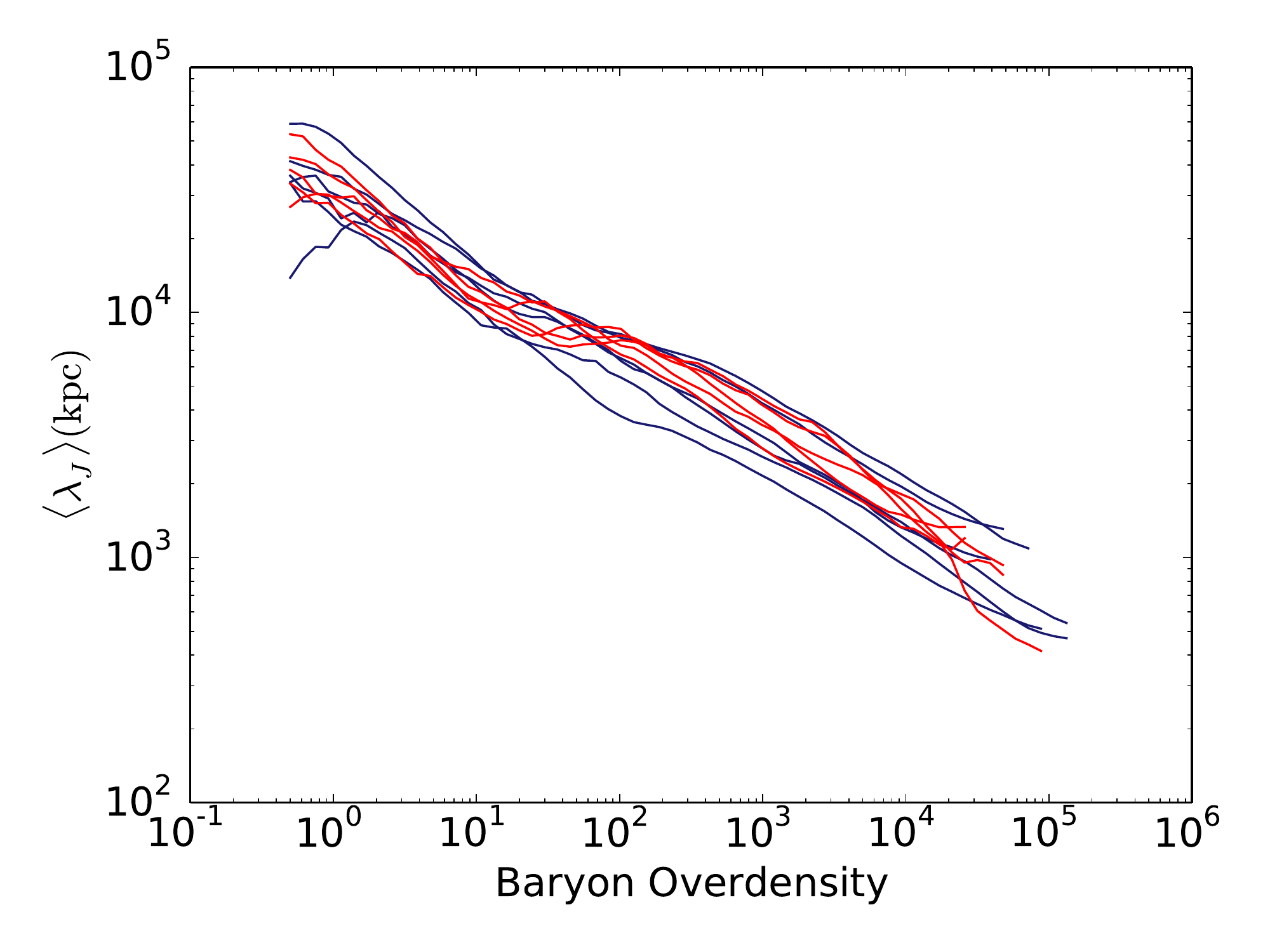}
  \includegraphics[width=0.45\textwidth]{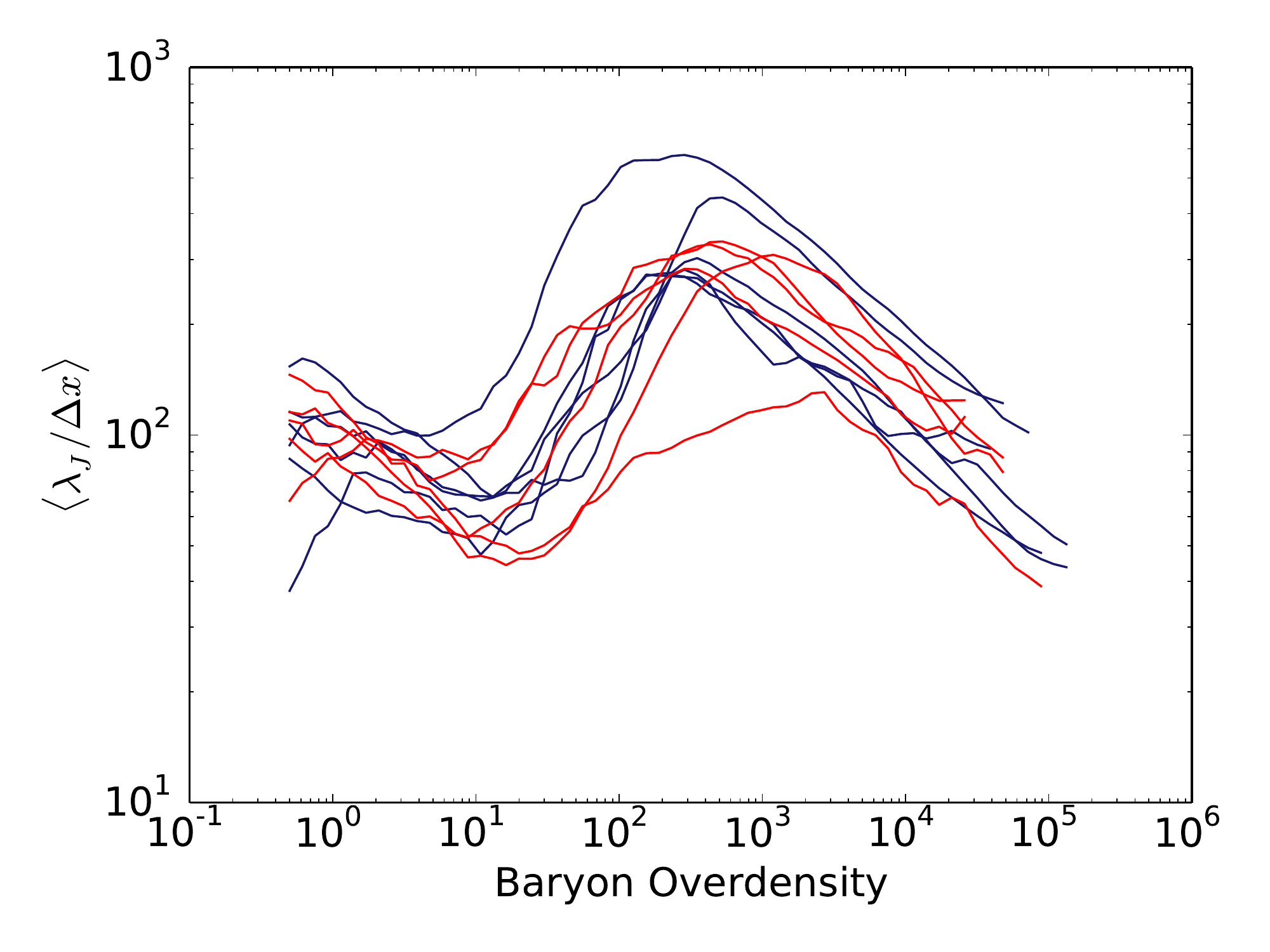}
  \caption{Volume-averaged Jeans length versus baryon
           overdensity for relaxed and unrelaxed clusters. Red
           signifies unrelaxed clusters while blue indicates relaxed,
           with one line per cluster. \emph{Left:} Mean value of
           Jeans length, \emph{Right:} Mean value of Jeans length
           divided by cell size.}
\label{fig:jeans_length}
\end{figure*}

The Jeans length

\begin{equation}
\lambda_J = \sqrt{\frac{15k_BT}{4\pi G\mu\rho}}
\end{equation}
is the critical length below which a self-gravitating cloud of gas will collapse.
Figure~\ref{fig:jeans_length} shows the mean Jeans length as
a function of baryon overdensity, and the mean cell size normalized
Jeans length. As these simulations are adiabatic, the the direct
$\lambda_J$-$\rho$ relation is a straightforward power law; however,
as the adaptive mesh does not resolve every cell in the cluster to the
highest level, the $\lambda_J/\Delta x$-$\rho$ relation is not
an exact power law. Despite
this, we always resolve the local Jeans length by at least 40 grid cells, 
and occasionally resolving it by up to a few hundred.

\citet{2011ApJ...731...62F} find that the Jeans length must be 
resolved by at least 30 cells to see dynamo action, and without increased 
resolution most of the amplification will still be due to compressive forces. 
Our simulations do exceed this minimum threshold so we may expect to see
some dynamo action, but much of the amplification may still be due to 
compression. 

\subsubsection{Resistive Length}

\begin{figure*}
  \centering
  \includegraphics[width=0.45\textwidth]{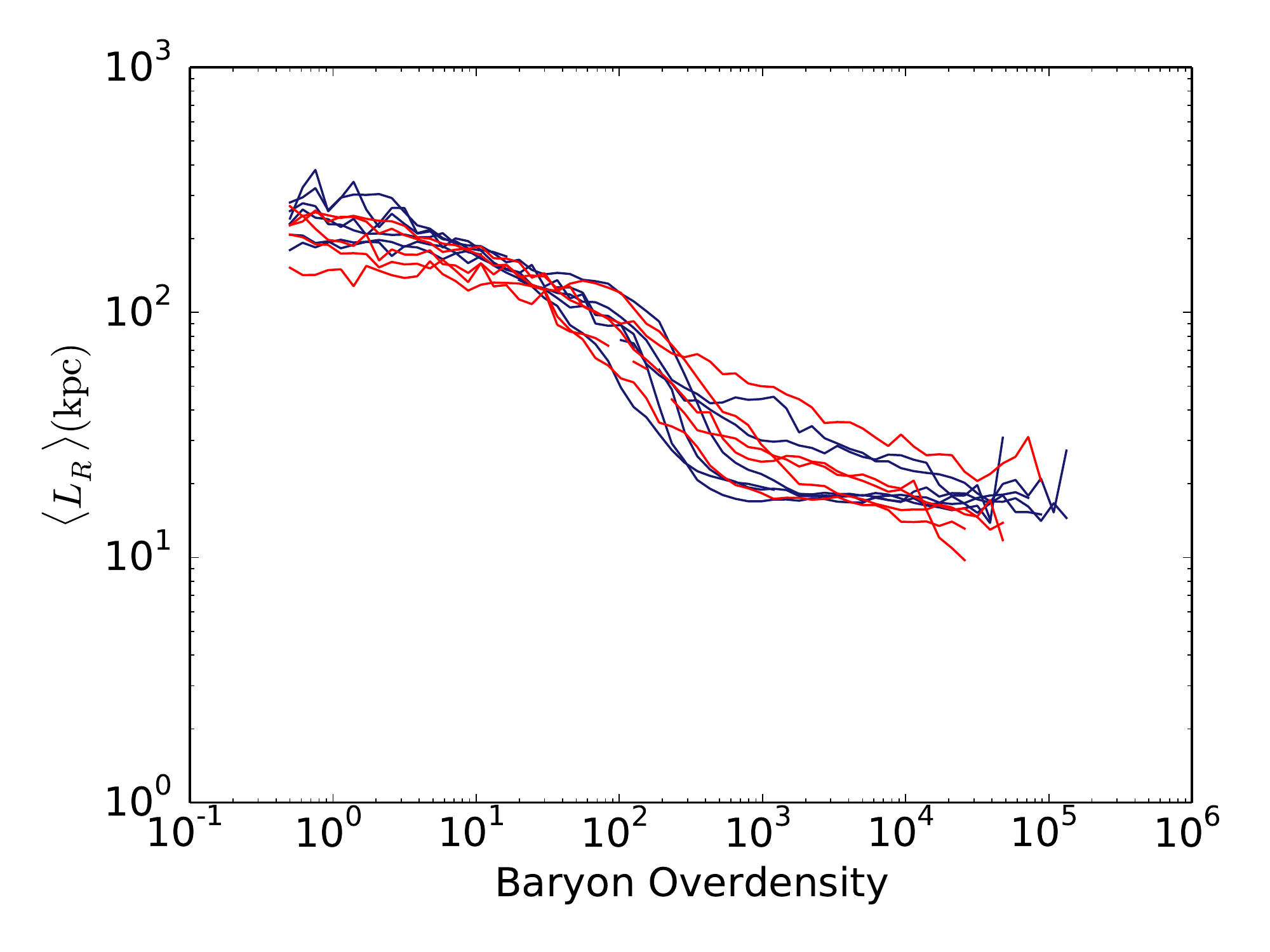}
  \includegraphics[width=0.45\textwidth]{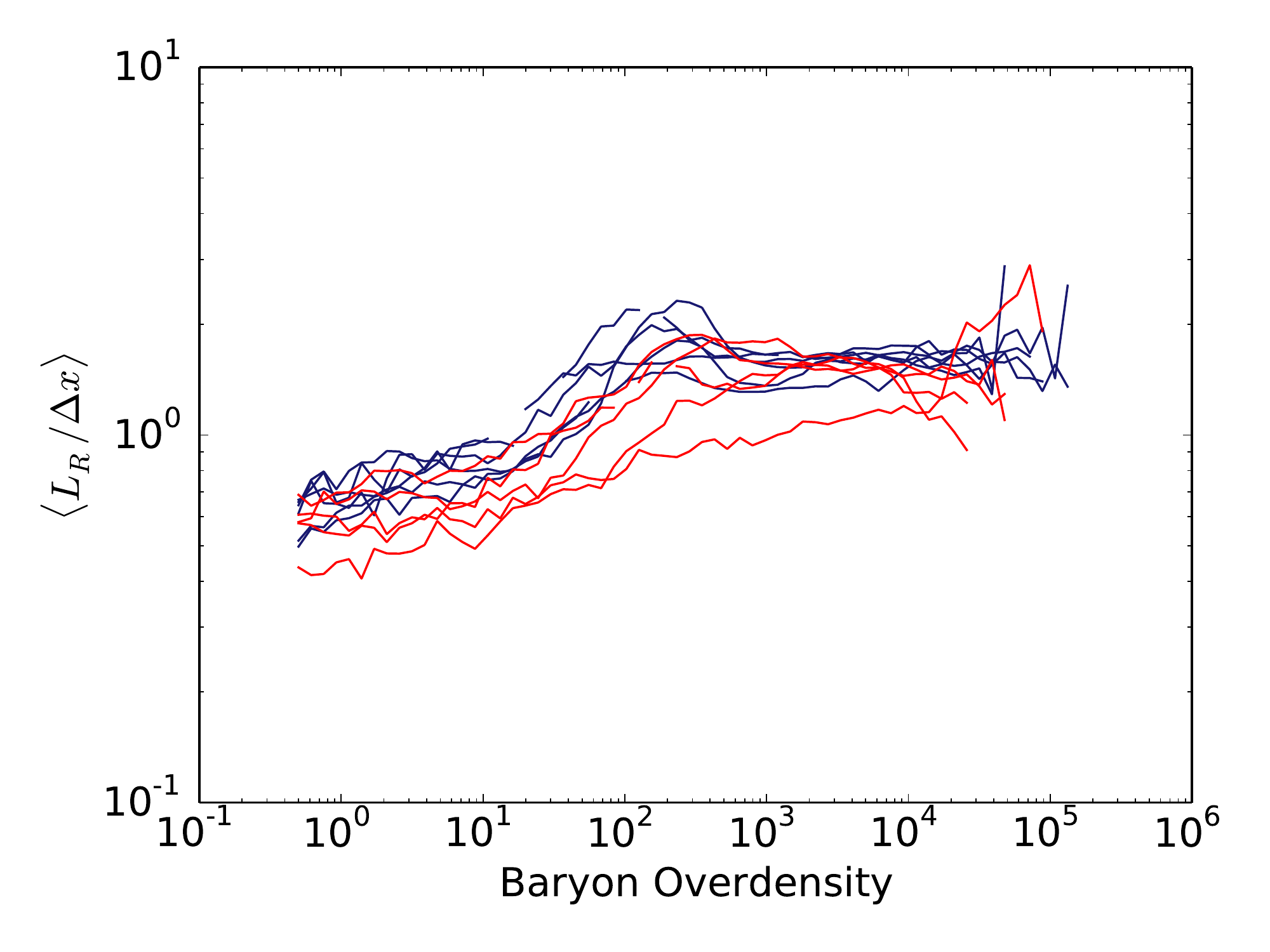}
  \caption{Volume-averaged characteristic resistive length versus baryon
           overdensity for relaxed and unrelaxed clusters. Red
           signifies unrelaxed clusters while blue indicates relaxed,
           with one line per cluster. \emph{Left:} Mean value of
           resistive length, \emph{Right:} Mean value of resistive length
           divided by cell size.}
\label{fig:resistive_length}
\end{figure*}

As mentioned in Section~\ref{sec:methods}, these simulations were refined
on both a density and a resistive length based criterion. In 
Figure~\ref{fig:resistive_length} we show the mean characteristic
resistive length 

\begin{equation}
  L_R = |B|/|\nabla \times B|
\end{equation}

and the resistive length divided by the cell size as a function of baryon 
overdensity. As discussed in \citet{2011ApJ...739...77X}, it was critical
to also refine based on resistive length in order to achieve even modest
levels of magnetic field amplification. 

Despite a wide variety of cluster relaxation states and physical conditions, the clusters have
very similar resistive lengths as a function of baryon overdensity. In
general, there are values of $\sim100$~kpc at cluster outskirts and
values of $\sim10$~kpc at cluster centers, with steep transition happening
around baryon overdensities of $10^2-10^3$ (radii of $0.1-0.5~R_{vir}$)
for most of clusters. As enforced by the refinement criterion,
$L_R/(\Delta x)$ maxes out for most of the clusters at just over one.

Note that the resistive MHD equations are not actually being solved here;
$L_R$ is the "characteristic" local resistive length, and the actual 
resistive dissipation scale will be much smaller. The sharp floor of
the resistive length function is indicative of the limitations in 
simulation spatial resolution, not of a physical process. This gives some indication
of where in the cluster we will see the effects of a finite numerical
resistivity directly affecting the magnetic field properties.

\subsubsection{Correlation Length}
\label{sec:autocor}
Rotation measure observations find that the cluster magnetic field is
patchy, and turbulent down to scales of $10$~kpc - $100$~pc
\citep{1995A&A...302..680F,2010arXiv1009.1233B}. Additionally,
many observations are interpreted in light of a single scale model where
the magnetic field is assumed to be composed of uniform cells of size
$\Lambda_c$ with random orientation \citep{1982ApJ...252...81L}. This 
produces a Gaussian distribution of patchy magnetic field with zero mean
and dispersion $\sigma_{RM}$. \citet{2004A&A...424..429M} find that using the 
autocorrelation length as a proxy for $\Lambda_c$ produces the closest
matching $\sigma_{RM}$ profiles. As our simulations have a maximum
resolution of $10$~kpc, it is unlikely that field alignment is driven
by the same processes at small scales. As one means of addressing the
magnetic field spatial distribution in our simulated clusters, we
examine the magnetic field autocorrelation length in the intracluster 
medium. The autocorrelation length is defined as

  \begin{equation}
    \label{eq:autocor}
    \lambda_{B_z} = \frac{\int_0^\infty <B_z(\vec x) B_z(\vec x +
       \vec{\ell})>d\ell}{<B_z(r)^2>} 
  \end{equation}

To calculate the autocorrelation function, pairs of points were picked such 
that the distance between them falls within a given range of $\ell$. We 
then find the average of $B_z(\vec{x})B_z(\vec{x}+\vec{\ell})$ for all
the pairs of points within 
the bin. The results are insensitive to the choice of magnetic field
orientation used; for brevity, we show only $B_z$. 

\begin{figure*}
  \label{fig:autocor_shells_mag}
  \centering
  \includegraphics[width=0.95\textwidth]{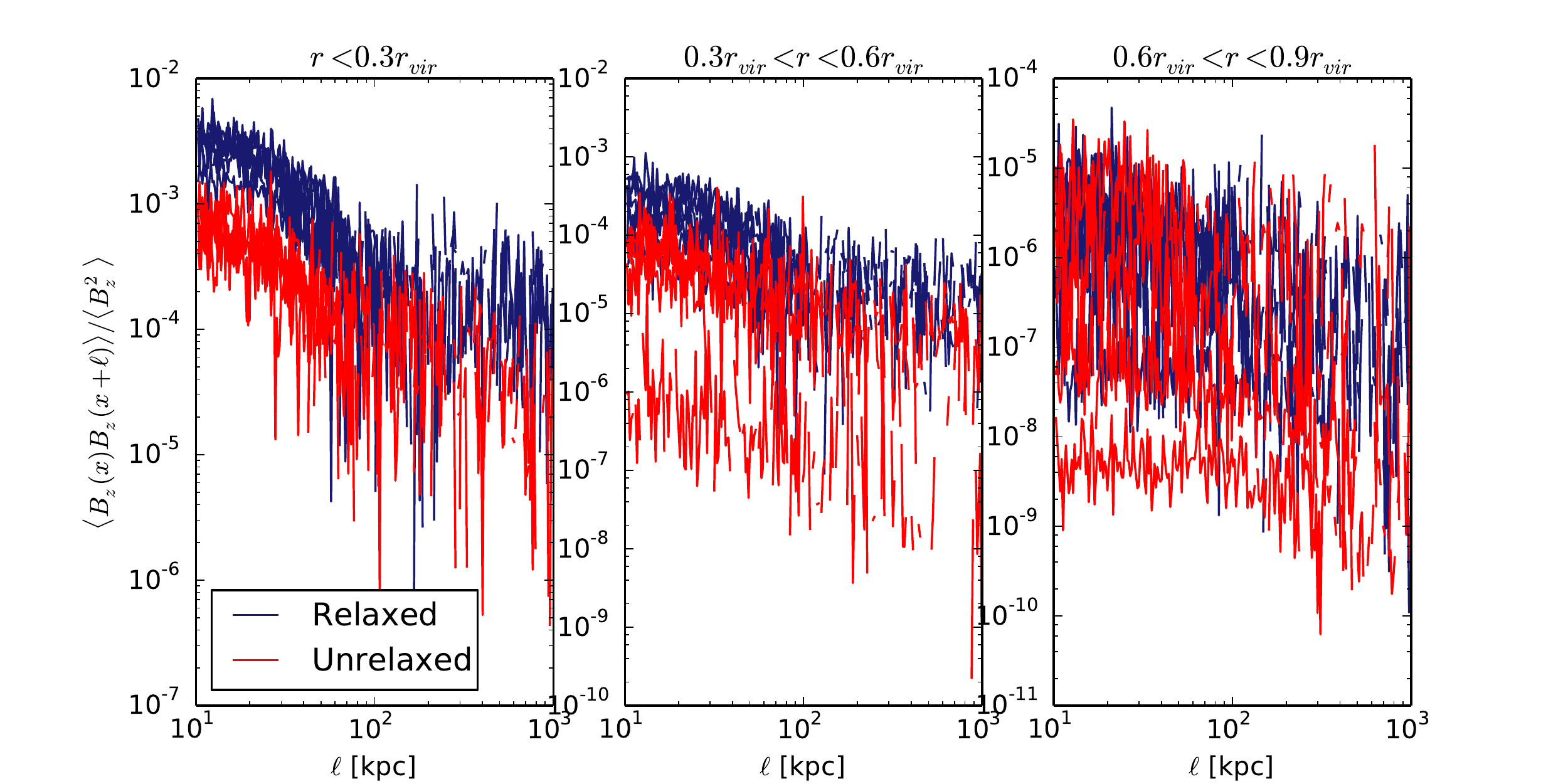}
  \caption{Magnetic field autocorrelation function calculated in
           spherical shells. Each panel shows one spherical shell with a line
           corresponding to a single cluster. Blue lines indicate relaxed
           clusters, while red lines indicate unrelaxed clusters. Higher 
           numbers indicate a greater degree of correlation.}
  \vspace{8mm}
\end{figure*}

A global autocorrelation function is of limited use due to its contamination
by phenomena such as structure in large scale density distributions, the
cluster gravitational potential, and bulk flows. As such we show the 
autocorrelation function sperately for a a several spherical shells through
the cluster. Figure~\ref{fig:autocor_shells_mag} shows the autocorrelation 
functions plotted for every cluster with each shell in a different panel, and 
the colors indicating relaxation state. We find that the magnetic field is more 
correlated closer to the cluster center; however, cells are not guaranteed 
to be at maximum resolution at the outskirts of the cluster which may influence
the observed autocorrelation. After normalizing for average magnetic field
magnitude, relaxed clusters have greater degrees of autocorrelation than
unrelaxed clusters, particularly in cluster centers. The turnover occurs at 
roughly 80~kpc for both relaxed and unrelaxed clusters.

  \subsection{Energetics}
Here we discuss energy and energy transfer as a function of 
scale; both of these measures can provide key insights into
the processes acting to amplify the magnetic field.

\subsubsection{Structure Functions}
\label{sec:struc}
Magnetic field amplification processes will impart structure on
the magnetic field power spectra, particularly in the inertial range,
while numerical effects are likely to dominate at small scales.
As in Section \label{sec:autocor} we work around this limitation 
imposed by global cluster structure by  examining structure
functions of the kinetic energy and magnetic field in spherical shells.
The structure function of a quantity $A(\vec x)$ of order $p$ takes
the form

\begin{equation}
  S_p(l) = <|A(\vec x)-A(\vec x + \vec{dx})|^p>
\end{equation}

We look at exclusively functions of order $p=2$ because a power spectrum 
with scaling index $\alpha$ has a second order structure function with scaling
index $l=-\alpha-1$. For pure Kolmogorov turbulence this corresponds to
$\alpha=-5/3$ and $l=2/3$.

To calculate the structure functions we randomly selected pairs of points
with separation $l$, uniformly distributed within a thick spherical
shell centered on the densest cell in the simulated cluster.
The quantity $|A(\vec x)-A(\vec x + l)|^2$ is then calculated for each point
pair and is averaged over bins in $l$. 62,500 point pairs were found for
250 $l$ bins with 250 points per bin. The total number of points was
chosen to be large but not oversample the central region in any
cluster.  We note that, within reason, the result is robust to the choice of number of
points and bins.

\begin{figure*}
  \label{fig:struc_shells_momentum}
  \centering
  \includegraphics[width=0.95\textwidth]{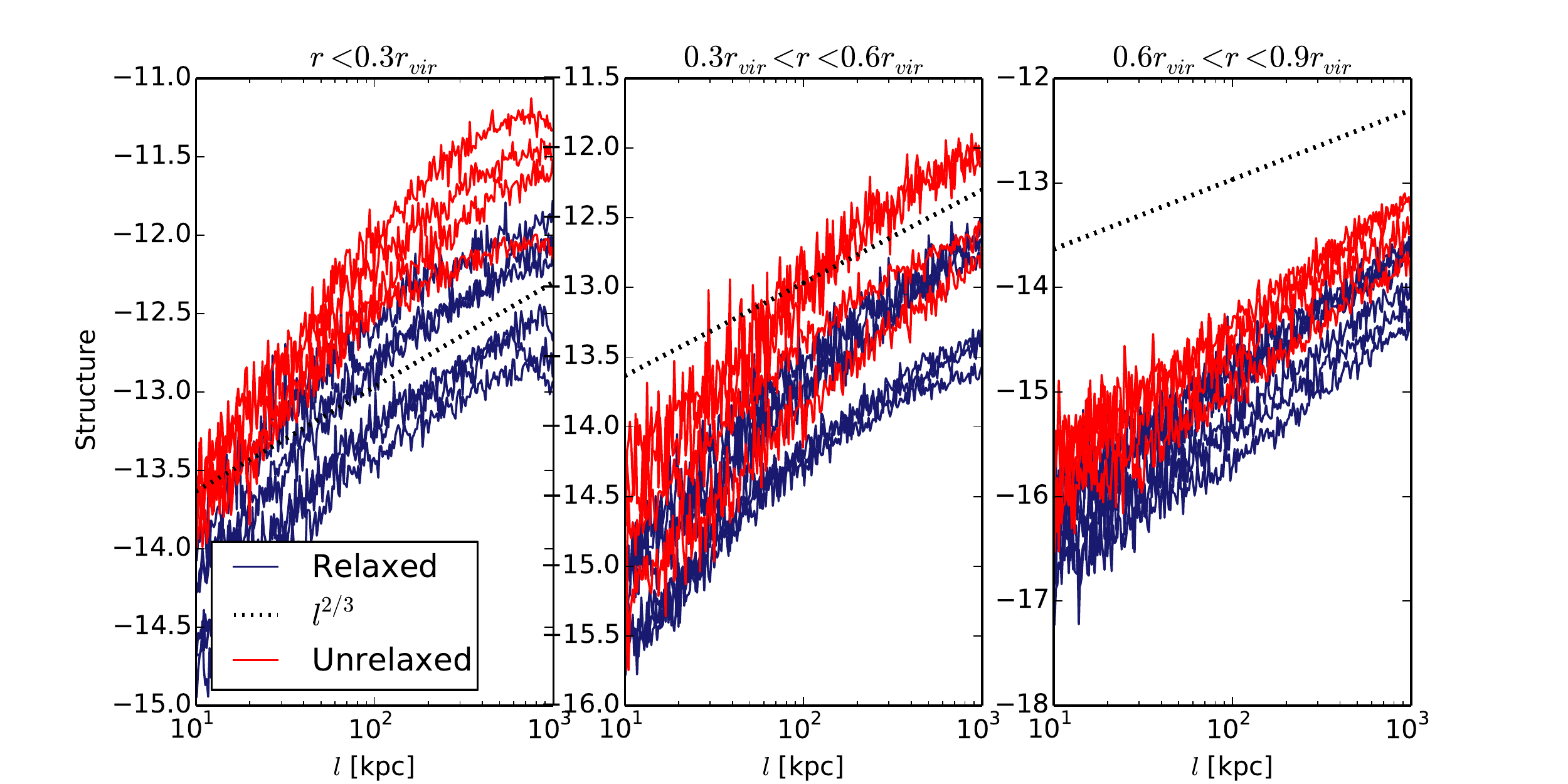}
  \caption{Smoothed momentum structure function calculated in spherical shells. Each panel shows one spherical shell
           with a line corresponding to a single cluster. Blue lines indicate relaxed clusters,
           while red lines indicate unrelaxed clusters.  The dashed line indicates a power law
           of $S_2(\rho u)\sim l^{4/3}$ and the dotted line indicates a power law of
           $S_2(\rho u)\sim l^{2/3}$ (Kolmogorov turbulence). }
  \vspace{8mm}
\end{figure*}

In Figure~\ref{fig:struc_shells_momentum} we plot the momentum structure
function for each of the three spherical shells. At the center of the 
clusters ($r<0.3 R_{vir}$) nearly all clusters show some inertial range
from $\sim 80-900$ kpc. The inertial range has a slope of $S_2\sim l^{2/3}$,
typical of incompressible Kolmogorov turbulence. In the middle shell 
($0.3R_{vir}<r<0.6R_{vir}$) only a few of the more massive clusters show
any sort of inertial range. At the largest radii ($0.6R_{vir}<r<0.9 R_{vir}$)
no clusters show any sort of inertial range. 
It is also possible that the compressible nature of the turbulence makes the 
scaling behave more like supersonic turbulence, in which case one would expect a
steeper slope in the inertial range \citep{2007ApJ...665..416K}
We note that the smaller
wavenumbers ($l< 30 \Delta x = 300$ kpc) may be contaminated by 
numerical effects \citep{2009A&A...508..541K}. 

\begin{figure*}
  \label{fig:struc_shells_mag}
  \centering
  \includegraphics[width=0.95\textwidth]{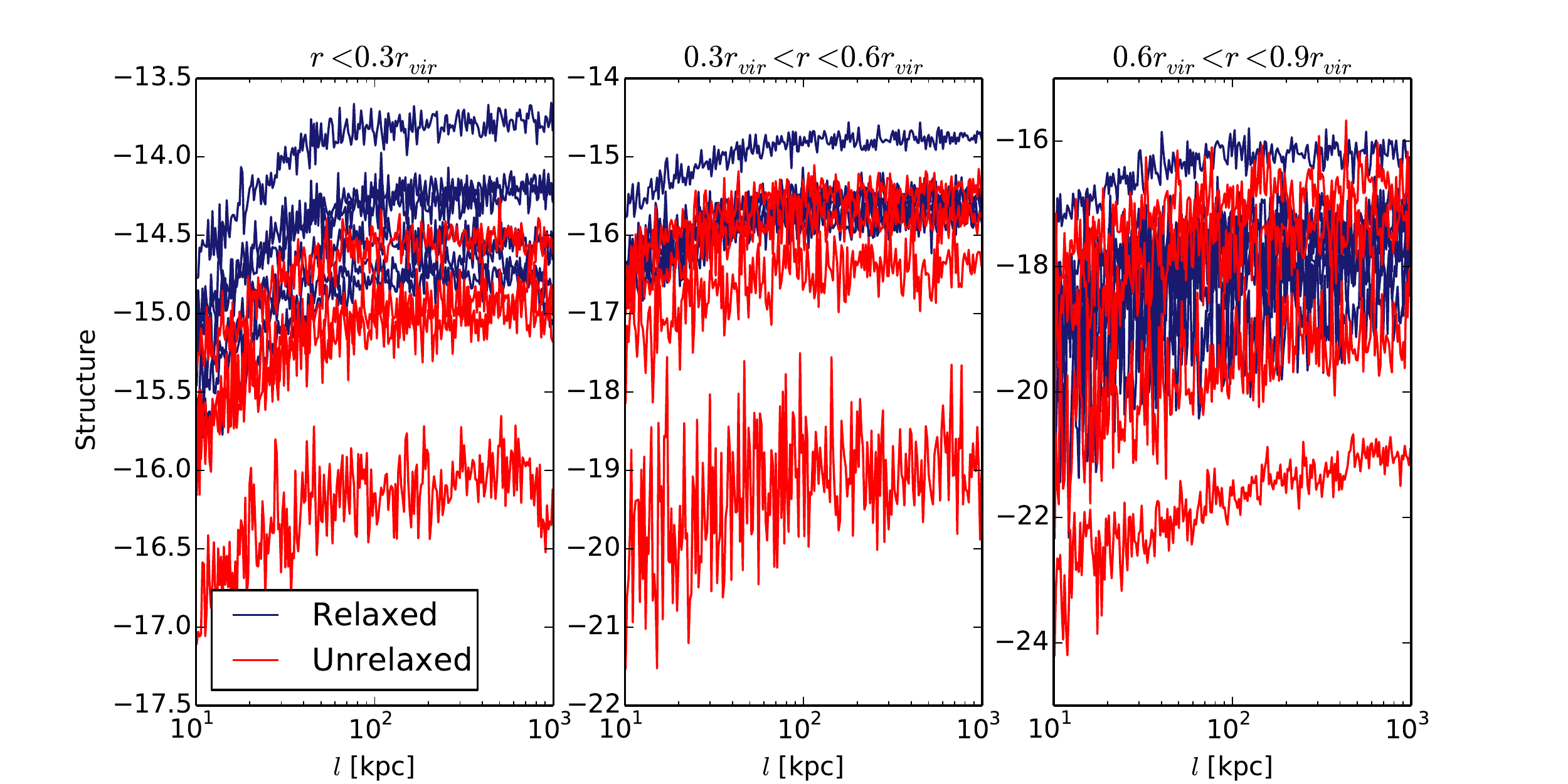}
  \caption{Smoothed magnetic field structure function calculated in spherical shells. Each panel shows one spherical shell
           with a line corresponding to a single cluster. Blue lines indicate relaxed clusters,
           while red lines indicate unrelaxed clusters. }
  \vspace{8mm}
\end{figure*}

Figure~\ref{fig:struc_shells_mag} shows the magnetic field structure
function for the three spherical shells. Here the spectra across 
most radii show a slope of $S_2\sim l^0$, with a microscale turnover at
roughly 60 kpc. As per the magnetic energy plots in 
Figure~\ref{fig:prof_mag_std}, the relaxed clusters have more magnetic 
energy than the unrelaxed clusters. As discussed by 
\citet{2011ApJ...739...77X}, 
the total magnetic energy is less than the total kinetic energy. 
This also holds for scale by scale energy partition; unlike saturated 
smale-scale dynamo (SSD) predictions, there is no scale below which magnetic energy is in 
equipartition with the kinetic energy at any radius.

\subsubsection{Spectral Energy Transfer Functions}
\label{sec:shellshell}
Spectral energy transfer analysis was a technique developed
by \citet{1967PhFl...10.1417K} to probe the methods by which
energy transfer occurs between energy reservoirs, and is one of
the most direct methods for examining the amplification and
dissipation processes acting on the magnetic field. A power spectrum
of the form $E(k)$ describes the total energy distribution as a 
function of wave number, or spatial scale, whereas a transfer 
function spectrum shows the total transfer of energy into or out
of scale $k$ of some energy reservoir due to a force-mediated
interaction with another reservoir. We give a detailed derivation
of the transfer function values in the Appendix. This closely
follows the approach of \citet{2010ApJ...714.1606P}, but is extended
to include self-gravitation.

In general the transfer functions are denoted $T_{XYZ}(k)$, which
indicates transfer from energy reservoir $X$, to scale $k$ of
reservoir $Y$, via force $Z$. In this analysis, the source reservoir is 
integrated over all wavenumbers, while the destination is at a specific
wavenumber, $k$ The transfer function values are
calculated by taking a small ($100^3$ cells, $\sim 1$ Mpc), fixed resolution
cube at the center of the cluster, taking the appropriate quantities'
Fourier transforms and dotting them together accordingly, as per
the discussion in the Appendix. We chose to only use this small
central box because clusters have strong radial dependence. By using
a small cube in the center we approximate an idealized turbulent
box. 


More specifically, we look at the transfer functions $T_{KBT}$,
$T_{BKT}$, $T_{KBP}$, and $T_{BKP}$. $T_{KBT}$ (and $T_{BKT}$) measure
the transfer of kinetic energy to magnetic energy (and vice versa) due
to tension forces. By tension forces we mean energy transfer due to
stretching and bending of field lines. $T_{KBP}$ and $T_{BKP}$ are
the energy transfers due to magnetic pressure and the internal
magnetic cascade. In a compressible fluid these two terms
cannot be disentangled from each other \textit{a priori}, whereas an incompressible
fluid guarantees that $\nabla\cdot\vec u=0$, making the 
transfer all due to internal magnetic cascade \citep{2010ApJ...714.1606P}.

\begin{figure}
  \label{fig:spect_KBT}
  \centering
  \includegraphics[width=0.45\textwidth]{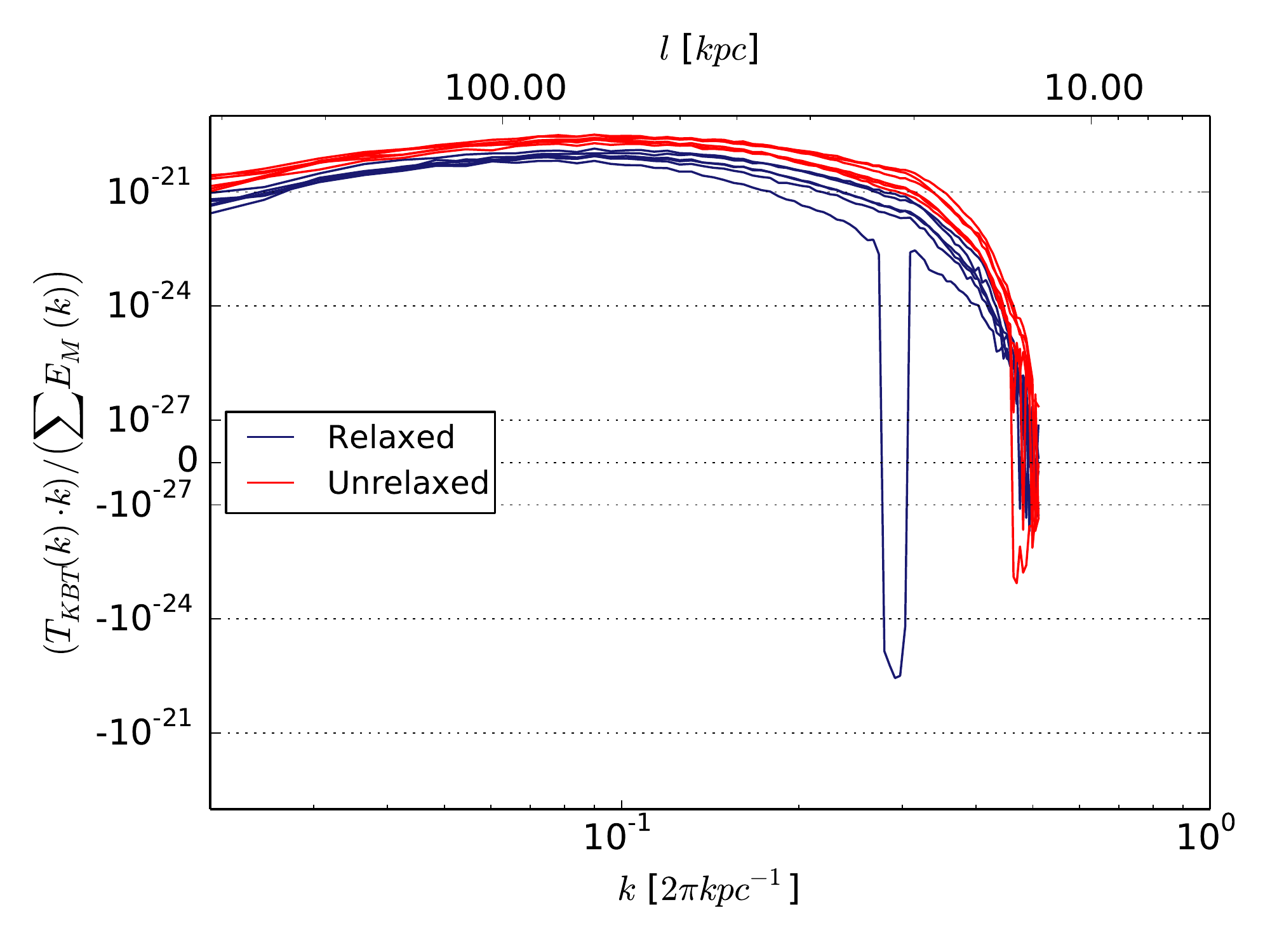}
  \caption{Transfer of kinetic energy of all scales to magnetic energy at
    scale $k$, via tension, normalized by magnetic energy. Positive
    values indicate kinetic energy is transforming to
    magnetic energy at scale $k$, while negative values indicate
    the magnetic energy at scale $k$ is losing energy to the kinetic
    energy reservoir. Relaxed clusters are in \emph{blue} and 
    unrelaxed clusters are in \emph{red}. The large dip is present only
    in cluster 3.}
  \vspace{8mm}
\end{figure}

Figure \ref{fig:spect_KBT} shows transfer of energy from the kinetic
energy reservoir to scale $k$ of the magnetic energy reservoir due
to the tension force. As the values are nearly all positive, every
scale of magnetic field has energy being transfered from kinetic energy.
Additionally, the magnitude of the normalized energy transfer is very
similar across all clusters, relaxed and unrelaxed. There is one
relaxed cluster that has some energy transfer away from the magnetic
energy at a scale of around 20~kpc.

\begin{figure}
  \label{fig:spect_BKT}
  \centering
  \includegraphics[width=0.45\textwidth]{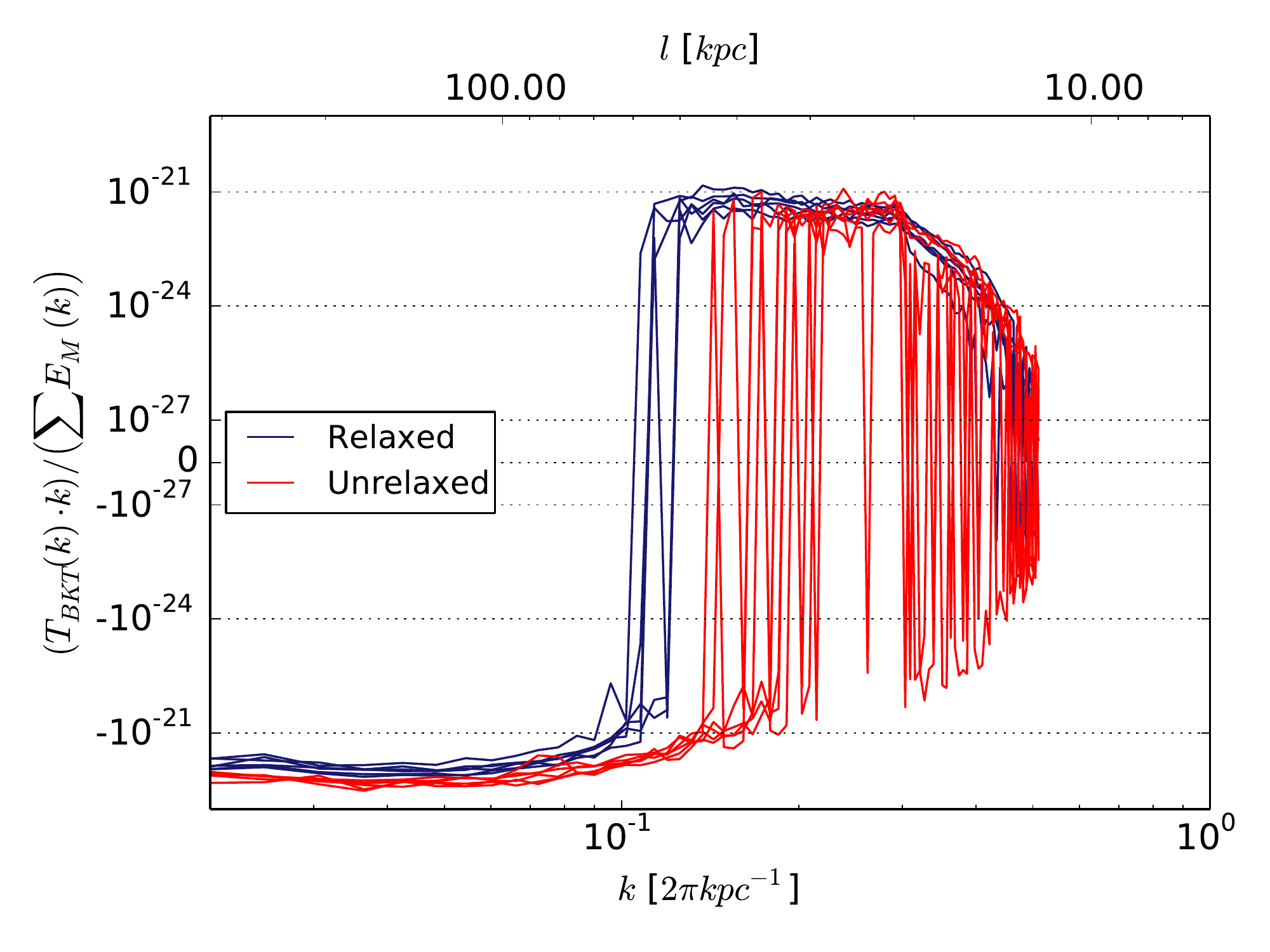}
  \caption{Transfer of magnetic energy of all scales to kinetic energy at
    scale $k$, via tension, normalized by magnetic energy. Positive
    values indicate magnetic energy is transforming to
    kinetic energy at scale $k$, while negative values indicate
    the kinetic energy at scale $k$ is losing energy to the magntic
    energy reservoir. Relaxed clusters are in \emph{blue} and 
    unrelaxed clusters are in \emph{red}.}
  \vspace{8mm}
\end{figure}

Figure \ref{fig:spect_BKT} is the transfer of energy from the magnetic
energy reservoir to scale $k$ of the kinetic energy. In general, the
larger scales of kinetic energy lose energy to magnetic fields. In
relaxed clusters and some unrelaxed clusters the smaller scales of
kinetic energy take energy from the magnetic field. In other unrelaxed
clusters the transfer function is very noisy at smaller scales,
indicating that neither transfer direction dominates (i.e., comparable
amounts of energy are transferred from magnetic to kinetic energy, and
vice versa, at small scales). The turnover
between energy transfer from and energy transfer to magnetic fields appears to occur
at a slightly larger scale for relaxed clusters than unrelaxed clusters,
but the sample size is quite small. In general this happens around
$40-60$~kpc. 

\begin{figure}
  \label{fig:spect_KBP}
  \centering
  \includegraphics[width=0.45\textwidth]{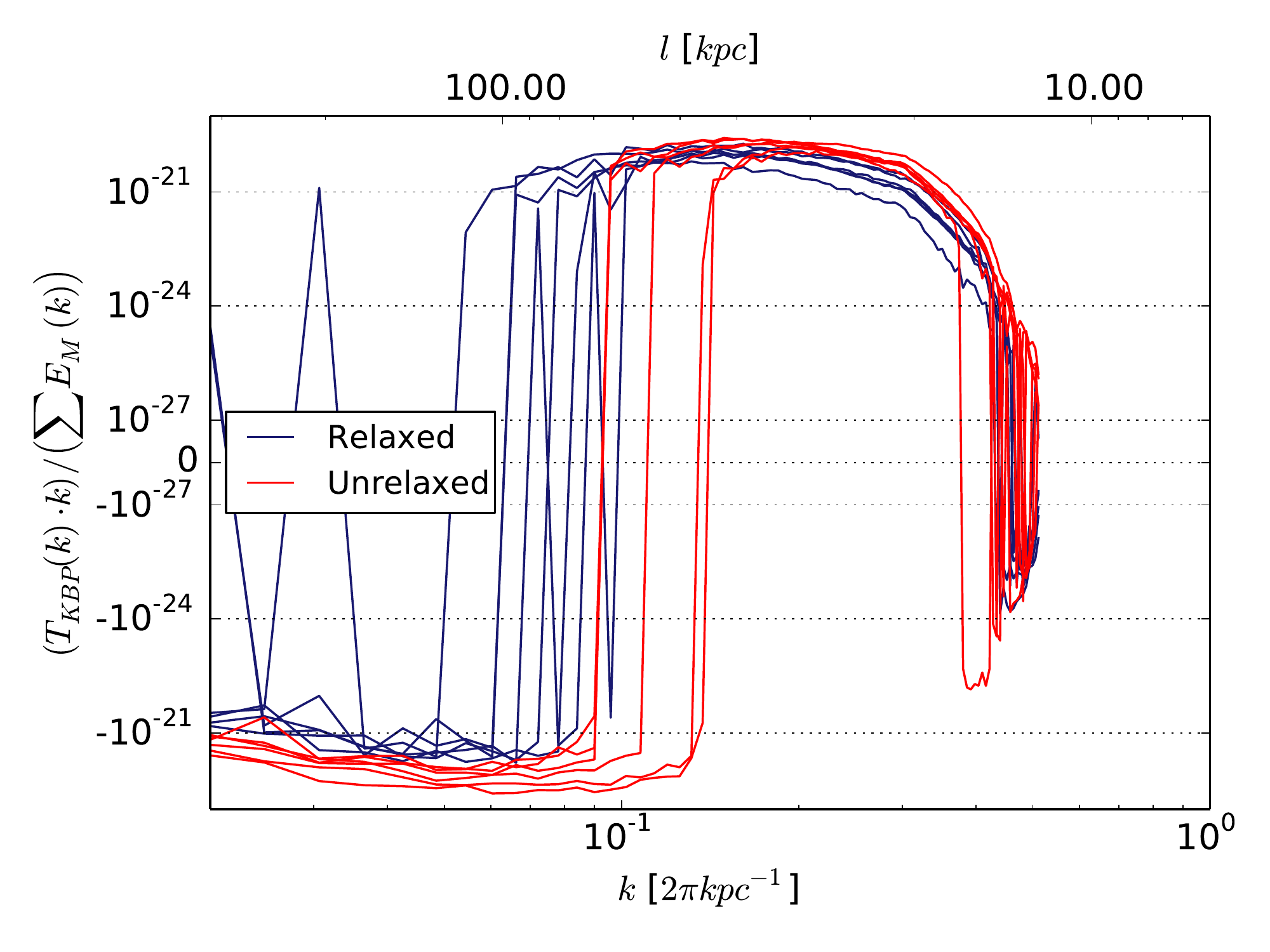}
  \caption{Transfer of kinetic energy of all scales to magnetic energy at
    scale $k$, via pressure, normalized by magnetic energy. Positive
    values indicate kinetic energy is transforming to
    magnetic energy at scale $k$, while negative values indicate
    the magnetic energy at scale $k$ is losing energy to the kinetic
    energy reservoir. Relaxed clusters are in \emph{blue} and 
    unrelaxed clusters are in \emph{red}.}
  \vspace{8mm}
\end{figure}

In Figure~\ref{fig:spect_KBP} we show the transfer of kinetic
energy to magnetic energy at scale $k$ via pressure. Generally, 
the smaller scales of magnetic energy gain energy from the kinetic
reservoir via pressure forces, while larger scales of magnetic
field show a less regular pattern but tend to lose energy to the
kinetic reservoir due to pressure. The point where the behavior
changes is widely dispersed between clusters but lies between scales
of 50 and 100 kpc.

We also examined the transfer of energy from the magnetic reservoir
to the kinetic reservoir at scale $k$ due to pressure, but found
that spectra for all clusters were exceedingly noisy. The magnitude
of normalized energy transfer remained the same for all clusters,
but oscillated between positive and negative values. This
indicates that the energy transfer due to this mechanism is in
rough equilibrium and does not have a large impact in the total
dynamics of the systems.

Because there is consistent loss of magnetic energy due to 
pressure at large scales and gain of magnetic energy due to 
pressure at small scales with no net gain or loss of kinetic
energy at any scale, this is consistent with compressible effects
being largely negligible and $T_{KBP}$ being equivalent to
the transfer function of the magnetic cascade.

\begin{figure}
\includegraphics[width=0.45\textwidth]{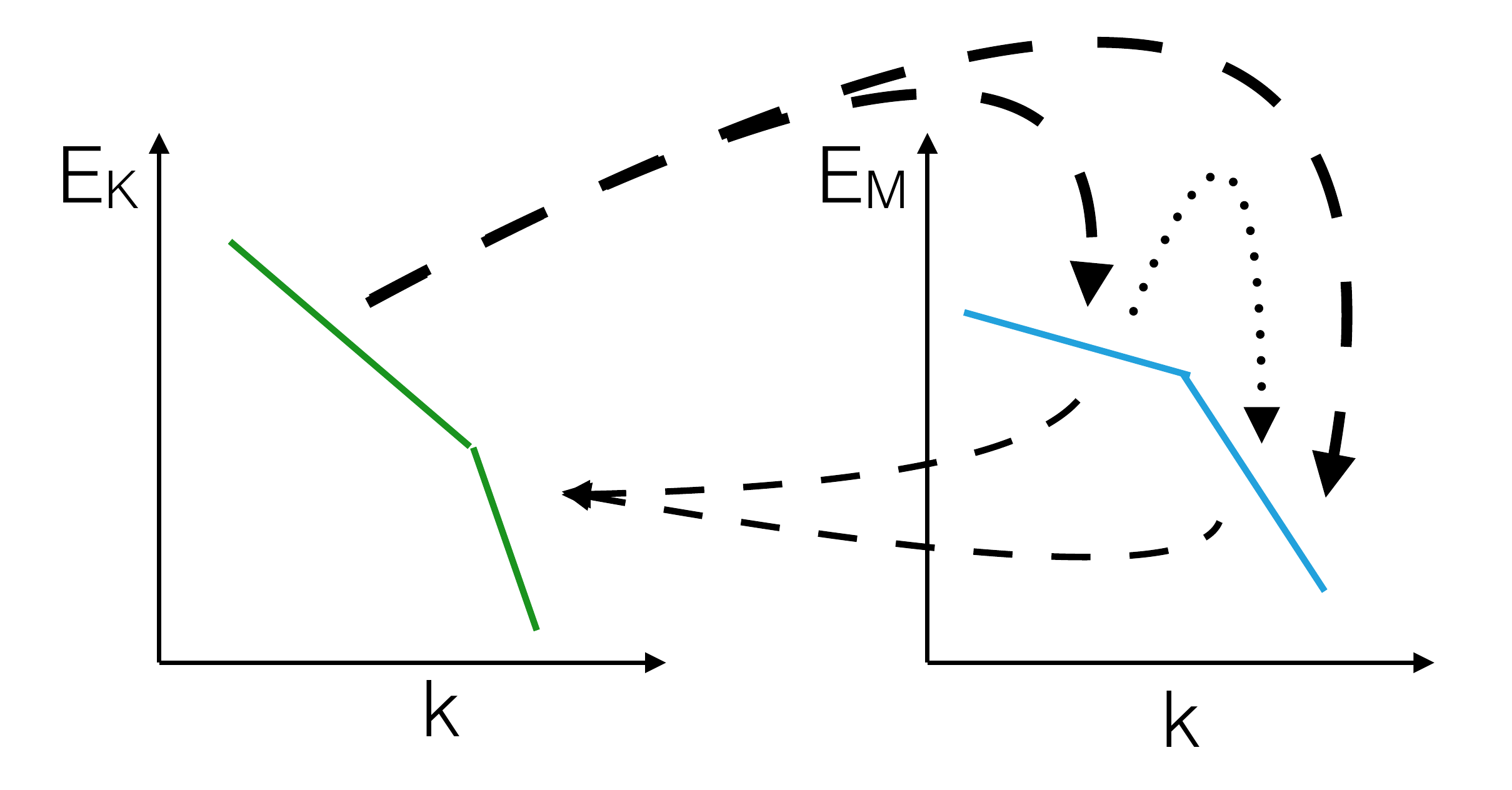}
\caption{Schematic example of the transfer between the kinetic
energy reservoir and the magnetic energy reservoir. Dashed
lines indicate transfer due to tension and dotted lines
indicate transfer due to compression.}
\label{fig:schem_transfer}
\end{figure}

These transfer functions depict a scenario where large scale fluid motion
bends the magnetic field at all scales, all scales of the magnetic
field do work to induce fluid motion at small scales via magnetic tension, and
compressive motions act to cascade energy from larger magnetic
field scales to smaller scales. This picture is shown schematically
in Figure~\ref{fig:schem_transfer}. 

\emph{This picture is consistent with a small scale dynamo stuck in 
the kinematic stage due to a lack of turbulent support at 
small scales.} \citet{2011ApJ...736...36M} show transfer functions for 
compressible simulations of small scale dynamo action in both the 
saturated and kinematic stage, and our results are qualitatively
similar to those in the kinematic stage. The magnetic field
is not able to grow to the saturated stage due to the energy
lost to the kinetic energy reservoir at small scales. As small
scale kinetic energy is dissipated by numerical viscosity, there
is less small scale turbulent motion to do work against magnetic
tension. Thus, the magnetic energy reservoir loses energy stored in
tension and is unable to build up and reach a saturated state
at small scales.

\section{Discussion}
\label{sec:discussion}

\begin{figure}
  \label{fig:schem_power}
  \centering
  \includegraphics[width=0.45\textwidth]{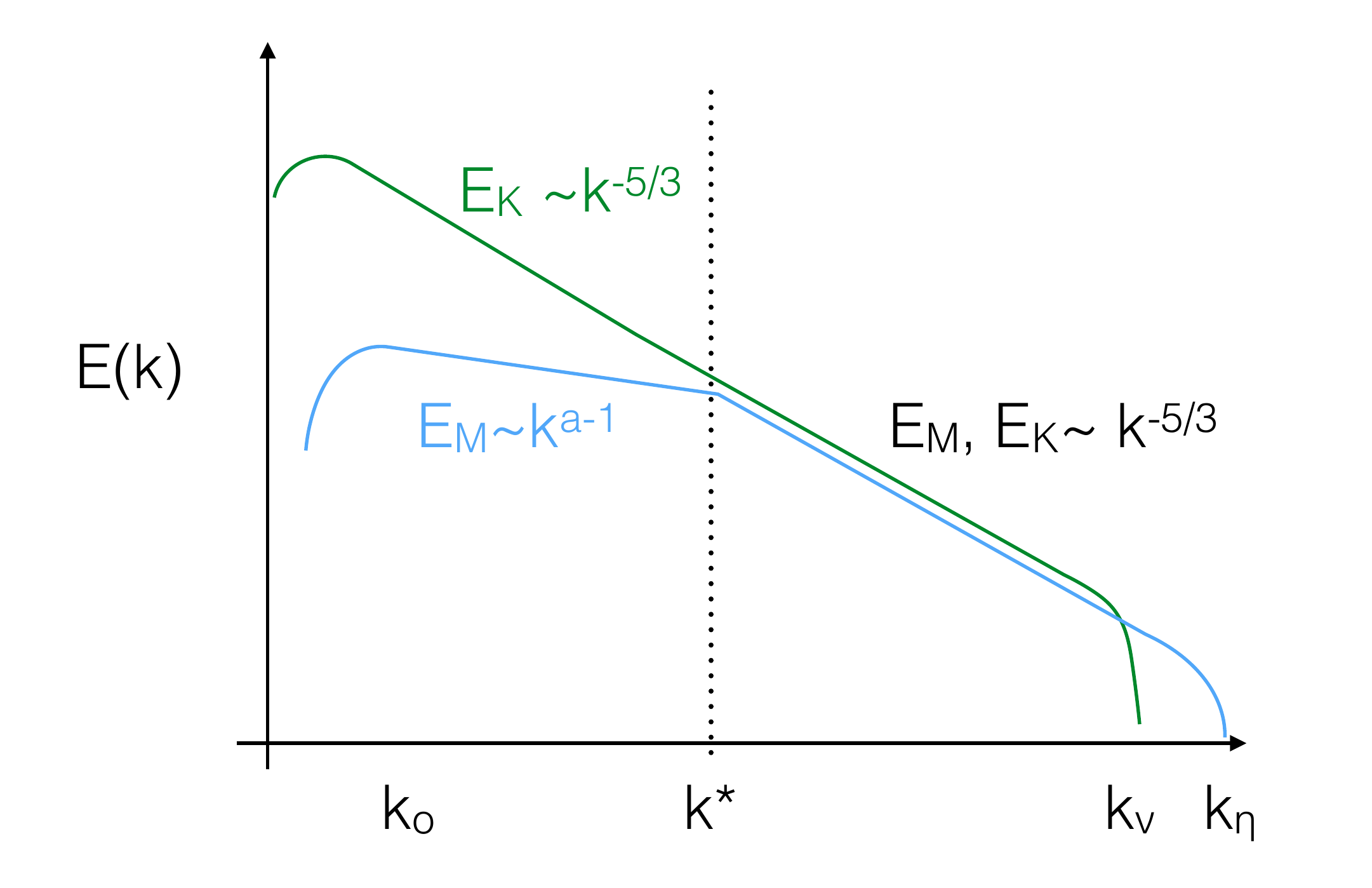}
  \caption{Schematic depiction of the magnetic and kinetic power
    spectra for a small scale dynamo. $k^*$ denotes the equipartition
    scale, $k_\nu$ is the viscous dissipation scale, and $k_\eta$
    is the resistive scale.}
  \vspace{8mm}
\end{figure}

Flux freezing and small-scale dynamo (SSD) action are two processes that can
drive the amplification of magnetic fields in clusters. A completely "frozen in"
magnetic field will  eventually be mixed across the entire volume due to fluid 
motion dragging the magnetic field. The resulting field 
will follow a $B\sim\rho^{2/3}$ power law. SSD action will act to bend and fold
a small, random initial magnetic field, quickly amplifying it. When the SSD is 
saturated the magnetic field energy will be in equipartition with the kinetic 
energy below some saturation scale, shown schematically in 
Figure~\ref{fig:schem_power}.

Both of these processes are active in the real intracluster medium.  A
high value of $\beta$ indicates that
the magnetic field is dynamically unimportant and thus completely
frozen in and thus the field will not impact the flow. A small-scale dynamo will
be active for large Reynolds and Prandtl numbers 
($Re\equiv UL/\nu$ and $Pm\equiv \nu/\eta$), both of
which  the ICM satisfies ($Re>10^{12}$, $Pr>10^{10}$). 

In simulations like the ones presented in this work, adiabatic expansion of the 
magnetic field from frozen-in field lines occurs, but true nonlinear
small-scale dynamo action
does not. Plasma $\beta$ values are very high in all regions of the cluster,
across many sizes and relaxtion states (as shown in
Figure~\ref{fig:phase_plasma_beta}). Reynolds  numbers and magnetic reynolds
numbers achieved by these simulations, however, are not nearly as high
as in the actual ICM
(in these calculations, $Re\sim 1000$, $Pr\sim1$).

The physical regimes where a small-scale dynamo can be active in has been the subject of recent debate.
\citet{2004MNRAS.353..947H} find that for $Pr\sim1$, the critical 
$R_{M}\sim 70$. \citet{2011ApJ...731...62F} find that the Jeans length must be 
resolved by at least 30 cells to see dynamo action, and without increased 
resolution most of the amplification will still be due to compressive forces. 
\citet{2012PhRvL.108c5002B} show that the transition to a nonlinear dynamo is 
strongly dependent on the effective Reynolds number; the timescale of the transition from 
the linear regime to the nonlinear regime is roughly 
$t_{linear} \sim t_{dyn} Re^{-1/2}$, and the magnetic energy growth rate is 
$\gamma\sim Re^{1/2}/30\tau_L$. Thus, for even reasonably large Re, the 
magnetic energy at time $t$ will be greatly reduced from expectations.

From estimates of $Re~(\sim1000)$ and $\Delta x/\lambda_J~(\sim 30)$, as well 
as the structure and transfer function analysis, it is clear that our
simulations do not have a nonlinear small-scale dynamo and are likely stuck in the linear 
phase. 

From \citet{2011ApJ...739...77X} we know that major mergers are a critical
driver of magnetic field amplification in these clusters. This is likely 
because the temporary increase in turbulence provides more kinetic energy 
which is available to be converted to magnetic energy. This transfer likely
takes the form of small scale kinetic energy doing work on the magnetic field 
to increase tension energy at all scales of the magnetic field.
Figure~\ref{fig:spect_KBT}  shows that only unrelaxed clusters with recent
mergers show any net flow from kinetic energy to small scales of magnetic
energy. All relaxed clusters have a net flow of small scale magnetic tension
energy to the kinetic reservoir.

That the turnover of the magnetic field structure function, $T_{KBT}$ 
transfer function and the end of the kinetic energy intertial range all occur 
at roughly 80 kpc ($\sim 8 \Delta x$) is not a coincidence. These metrics 
are all intertwined and critically depend on the numerical dissipation of the 
simulations. As \citet{2009A&A...508..541K} found similar limits to the 
inertial range due to numerical viscosity across a variety of codes and 
numerical methods, it is likely that a similar turnover range and magnetic 
field growth rate will be seen in other simulations.

This picture is likely worse in the outskirts of the cluster. Although
the AMR resolution criteria ensure refinement on both overdensity and
resistive length, the spatial resolution is poorer,  there is no inertial range
in the structure function, and there is remarkably little correlation
in the magnetic field. Although we cannot do the same transfer
function analysis in the outskirts due to the strong density gradients,
we expect that the transfer will be nearly entirely due to compressive
motions. Even if simulations were able to increase the spatial resolution in
the centers of clusters enough to ensure transition to the non-linear 
regime of dynamo action, the outskirts of the clusters would still lag
behind, likely  affecting global cluster properties.
 
Clearly it is not possible to excite ICM level magnetic field growth
via small-scale dynamo 
action in current simulations. Even the highest Reynolds numbers achievable by 
current generation simulations are orders of magnitude smaller than physically 
realistic values. Thus, the magnetic growth timescale will always be far too 
small. Increasing the simulation spatial resolution enough to diminish the effects of numerical 
viscosity is computationally unfeasible, so alternative models must be 
considered. \citet{2016ApJ...817..127B} consider a model where time-dependent 
global cluster turbulence analysis is used to derive the turbulent dissipation 
rate $\epsilon_{turb}$, which then constrains the possible magnetic energy 
density. Another possible approach could be to use a subgrid turbulence model, 
similar to that of \citet{2008ApJ...686..927S}, to add additional turbulent 
support and counteract the effects of numerical viscosity. Regardless, in order 
to study any process dependent on cluster magnetic fields, a model
that is more sophisticated than the current cosmological MHD
simulations must be utilized.

\section{Summary}
  \label{sec:summary}

In this paper we have further analyzed a set of 12 cosmological
magnetohdyrodynamic AMR simulations of galaxy clusters, originally presented by
\citet{2009ApJ...698L..14X,2010ApJ...725.2152X,2011ApJ...739...77X,2012ApJ...759...40X}.
The goal of this effort is to characterize the nature of the magnetized
intracluster plasma in order to inform more accurate plasma simulations, thus
guiding the creation of sub-grid models for plasma behavior that can be applied
to cosmological simulations.  Our primary results are as follows:

\begin{enumerate}

\item Although both flux-freezing and small-scale dynamo play a role
in the amplification of the magnetic field, most field amplification
is due to the compressive modes associated with flux-freezing.

\item Small-scale dynamo action is limited due to the low effective
Reynolds number of the ICM, which is small because of numerical
viscosity. This limits the inertial range of the kinetic turbulent
cascade, reducing the turbulent energy available to do work on the
magnetic field.

\item This picture is applicable across a variety of cluster relaxation states; 
although there is an increase in turbulence associated with major mergers, it 
is not enough to boost the small-scale dynamo out of the kinematic
stage.  Unrelaxed clusters 
 show more energy transfer from small scales of kinetic energy to magnetic 
energy than relaxed clusters, but it is only associated with a moderate increase
in magnetic field amplification.

\item We see even less small-scale dynamo action in the outskirts of
clusters as in the central region. As the spatial resolution is poorer
in the outskirts of the cluster, the Reynolds number is even higher,
making the timescale of the linear dynamo even longer.

\item Due to the similarity in the small-scale behavior of turbulent
cascades across a variety of astrophysical fluid codes
\citep{2009A&A...508..541K} and the strong dependence of small-scale
dynamo action on cluster turbulence \citep{2011ApJ...739...77X}, we
expect these results to be similar across a variety of simulation
codes.

\end{enumerate}

\acknowledgments
\section{Acknowledgments}

The authors would like to thank Kris Beckwith, Andrew Christlieb, Jeff
Oishi, and
Mark Voit for helpful discussions during the preparation of this
paper.  This work was supported by NASA through grants NNX12AC98G,
NNX15AP39G, 
Hubble Theory Grants HST-AR-13261.01-A and HST-AR-14315.001-A, and by the DOE Computational
Science Graduate Fellowship program.  The simulations presented in
this paper were performed on LANL supercoputing resources, and
analyzed on the TACC Stampede supercomputer under XSEDE allocations
TG-AST090040 and TG-AST100004. This work was supported in part by
Michigan State University through computational resources provided by
the Institute for Cyber-Enabled Research.  BWO was supported in part
by the sabbatical visitor program at the Michigan Institute for
Research in Astrophysics (MIRA) at the University of Michigan in Ann
Arbor, and gratefully acknowledges their hospitality. This work was also
supported by grants to J.O.B. from the National Science Foundation
(AST 1106437) and from NASA (NNX15AE17G). HL gratefully acknowledges
the support by LANL's LDRD program and DoE/OFES through CMSO. \texttt{Enzo}
and \texttt{yt} are developed by a large number of independent
researchers from numerous institutions around the world. Their
commitment to open science has helped make this work possible.


\bibliographystyle{apj}

\bibliography{apj-jour,ms}
\appendix
Spectral energy transfer functions were initially developed in the
incompressible limit by \citet{1967PhFl...10.1417K}. It was later extended to
include the compressible limit \citep{2007A&A...476.1113F,2009ApJ...690..974S,
2010ApJ...714.1606P}. Here we follow closely the approach of
\citet{2010ApJ...714.1606P} but extend the framework to include a
self-gravitating fluid. 

A galaxy cluster is a strongly stratified environment thhat is not generally
periodic, so we continue using a general orthonormal basis $\phi_k(\vvec{x})$
and arbitrary function $g(\vvec{x})$. Then our test function $g(\vvec{x})$ can be
written in k-space as

\begin{equation}
  \hhat{g}(k) = \int_\Omega g(\vvec{x})\phi_k(\vvec{x})d^3x
\end{equation}

with

\begin{equation}
  g(\vvec{x})=\sum_k\hhat g(k)\phi_k(\vvec{x})
\end{equation}

where $\Omega$ is the analysis volume. $\hhat g(k)$ is then the spectral density
of some global quantity $G$ in k-space, where $G$ is

\begin{equation}
  G = \int_\Omega g(\vvec{x})d^3 = \sum_k \hhat g(k).
\end{equation}

In real space the magnetic energy density can be written as $e_M = \frac{1}{8\pi}
|B|^2$. The total magnetic energy is

\begin{equation}
  E_M = \int_\Omega e_Mdx^3 = \sum_k E_M(k)
\end{equation}

so Parseval's theorem allows us to write the spectral energy density as

\begin{equation}
  E_M(k) = \frac{1}{8\pi} \hhat{\vvec{B}}(k) \cdot \hhat{\vvec{B}}^*(k).
\end{equation}
  
Similarly, the momentum energy density can be written $e_K=\frac{1}{2}\rho u^2$,
so the real kinetic spectral energy density can be written as

\begin{equation}
  \label{eq:EK}
  E_K(k) = \frac{1}{4}\left(\hhat{\vvec{u}}(k)\cdot\hhat{[\rho\vvec{u}]}\astc(k)
    + \hhat{[\rho\vvec{u}]}(k)\cdot\hhat{\vvec{u}}\astc(k)\right) 
\end{equation}

To break up the spectral energy densities into their components we write the
magnetic induction equation

\begin{equation}
\label{eq:induction}
  \frac{\partial \vvec{B}}{\partial t} = \nabla\times(\vvec{u}\times\vvec{B})
\end{equation}

and the conservative form of the momentum equation
\begin{equation}
\label{eq:consv_momentum}
  \frac{\partial(\rho\vvec{u})}{\partial t} + \nabla\cdot(\rho\vvec{uu}) =
  -\nabla P +(\nabla\times B)\times B-\rho\nabla\phi
\end{equation}

with $P$ as the thermal pressure and $\phi$ as the gravitational potential.
Combining the conservative form of the momentum equation with the 
continuity equation

\begin{equation}
  \frac{\partial\rho}{\partial t} + \nabla\cdot(\rho\vvec{u})=0
\end{equation}

allows us to write the momentum equation in terms of velocity

\begin{equation}
  \label{eq:prim_momentum}
  \frac{\partial\vvec{u}}{\partial t} = -(\vvec{u}\cdot\nabla)\vvec{u}-
  \frac{1}{\rho}\nabla P + \frac{1}{\rho}(\nabla\times B)\times B -\nabla\phi.
\end{equation}

The time evolution of the spectral magnetic energy can be written

\begin{equation}
  \frac{d E_M(k)}{dt} = \frac{1}{4\pi} \rm{Re}\left[\frac{\partial
  \hhat{\vvec{B}}}{\partial t}\cdot \hhat{\vvec{B}}\astc\right]
\end{equation}

so its component form can be derived by projecting the induction equation 
(Equation \ref{eq:induction}) onto its basis components in k-space, dotting
it with $\frac{1}{4\pi} \hhat{\vvec{B}\astc}(k)$, and taking the real component
to obtain

\begin{equation}
  \frac{d}{dt}E_M(k) = T_{KBP}+T_{KBT}+D_{B}
\end{equation} 

where
\begin{equation}
  T_{KBT}=\frac{1}{4\pi}\rm{Re}\left[ \hhat{\vvec{B}(k)}\cdot
    \hhat{[\vvec{B}\cdot\nabla\vvec{u}]\astc}(k)\right]
\end{equation}
represents the energy transfer rate from the kinetic energy reservoir to the
magnetic energy reservoir due to magnetic tension, 
\begin{equation}
  T_{KBP}=\frac{1}{4\pi}\rm{Re}\left[\hhat{\vvec{B}}(k)\cdot
    \hhat{[\vvec{B}\nabla\cdot\vvec{u}]\astc}(k)-\hhat{\vvec{B}}\cdot
    \hhat{[\vvec{u}\cdot\nabla\vvec{B}]\astc}(k)\right]
\end{equation}
represents the energy transfer rate from the kinetic reservoir to the magnetic
reservoir due to magnetic pressure, and $D_{B}$ is the dissapation due to
numerical resisitivity.

Similarly, the time evolution of the spectral kinetic energy can be written as

\begin{equation}
  \frac{d}{dt}E_K(k) = \frac{1}{2}\rm{Re}\left[ \hhat{\vvec{u}}(k)\cdot
  \frac{\partial \hhat{[\rho\vvec{u}]}\astc(k)}{\partial t} +
  \frac{\partial \hhat{\vvec{u}(k)}}{\partial t}\cdot
  \hhat{[\rho\vvec{u}]}\astc(k) \right]
\end{equation} 

so to break it into its components the conservative form of the momentum equation
(Equation \ref{eq:consv_momentum}) and the primitive form of the momentum
equation (Equation \ref{eq:prim_momentum}) are projected onto their basis
components and dotted with the appropriate vectors to find

\begin{equation}
  \frac{dE_K(k)}{dt} = T_{KKA}(k) + T_{KKC}(k) + T_{BK}(k) + T_{IK} + D_K
\end{equation}

Here, 

\begin{equation}
 T_{KKA}(k) = -\frac{1}{2} \rm{Re}\left[ \hhat{\vvec{u}}\cdot \hhat{[\vvec{u}
  \cdot \nabla (\rho \vvec{u})]}\astc(k)+\hhat{[\rho u]}\astc(k)\cdot\hhat{
  [\vvec{u}\cdot\nabla \vvec{u}]}(k)\right]
\end{equation}

is the transfer of kinetic energy inside the kinetic energy reservoir due to
advection,

\begin{equation}
  T_{KKC}(k) = -\frac{1}{2}\rm{Re}\left[\hhat{\vvec{u}}\cdot
  \hhat{[\rho\vvec{u}\nabla\cdot \vvec{u}]}\astc(k)\right]
\end{equation}

is the transfer of kinetic energy inside the kinetic reservoir due to 
compressible motion,

\begin{equation}
  T_{BKT} = \frac{1}{8\pi}\rm{Re}\left[\hhat{\vvec{u}}\cdot\hhat{[
  \vvec{B}\cdot\nabla\vvec{B}]}\astc(k)+\hhat{[\rho\vvec{u}]}\astc(k)
  \cdot\hhat{\left[\frac{1}{\rho}\vvec{B}\cdot\nabla\vvec{B}\right]}(k)
  \right]
\end{equation}

 is the transfer of energy from the magnetic energy reservoir due to
magnetic tension, and 

\begin{equation}
  T_{BKP} = \frac{1}{8\pi}\rm{Re}\left[\hhat{\vvec{u}}(k)\cdot\hhat{
  \left[-\frac{1}{2}\nabla|\vvec{B}|^2\right]}\astc(k) +\hhat{(
  \rho\vvec{u})}\astc(k) \cdot\hhat{\left[-\frac{1}{2\rho}\nabla
  |\vvec{B}|^2\right]}(k)\right]
\end{equation}

is the transfer of magnetic energy to kinetic energy due to pressure.

\end{document}